\documentclass[12pt]{iopart}

\pdfoutput=1

\usepackage{graphicx,sidecap}
\usepackage{bm}
\usepackage{braket}
\usepackage{amssymb}
\usepackage{amsfonts}
\usepackage{mathrsfs}
\usepackage{xcolor}
\usepackage{enumerate}
\usepackage[colorlinks,%bookmarks=false,
citecolor=blue,linkcolor=red,urlcolor=blue]{hyperref}

\newcommand\be{\begin{equation}}
\newcommand\bea{\begin{eqnarray}}
\newcommand\ee{\end{equation}}
\newcommand\eea{\end{eqnarray}}
\newcommand\nn{\nonumber\\}
\newcommand\vt{\tilde{v}}

\begin{document}

\title[Quantum quench within the gapless phase of the XXZ spin-chain]
{Quantum quench within the gapless phase of the spin-1/2 Heisenberg XXZ spin-chain}

\vspace{.5cm}

\author{Mario Collura$^1$, Pasquale Calabrese$^1$ and Fabian H. L. Essler$^{1,2}$}
\address{$^1$\,SISSA and INFN, via Bonomea 265, 34136 Trieste, Italy. \\}
\address{$^2$\,The Rudolf Peierls Centre for Theoretical Physics,
    Oxford University, Oxford, OX1 3NP, United Kingdom.}

%\date{\today}
\vspace{.5cm}

\begin{abstract}
We consider an interaction quench in the critical spin-1/2 Heisenberg
XXZ chain. We numerically compute the time evolution of the two-point
correlation functions of spin operators in the thermodynamic limit and
compare the results to predictions obtained in the framework
of the Luttinger liquid approximation. We find that the transverse
correlation function $\langle S^x_jS^x_{j+\ell}\rangle$ agrees with
the Luttinger model prediction to a surprising level of accuracy. The
agreement for the longitudinal two-point function 
$\langle S^z_jS^z_{j+\ell}\rangle$ is found to be much poorer.
We speculate that this difference between transverse and longitudinal
correlations has its origin in the locality properties of the
respective spin operator with respect to the underlying fermionic
modes. 
\end{abstract}

\maketitle

%\tableofcontents

\section{Introduction}
\label{sec intro}

The non-equilibrium dynamics of isolated quantum systems represents a
theoretical and experimental challenge that raises numerous fundamental
questions with applications to a variety of fields of modern physics. 
In a so-called \emph{interaction quench}, a system evolves unitarily
from an initial state, which is the ground-state of a translationally
invariant Hamiltonian differing from the one governing the time
evolution by an experimentally tunable interaction parameter
\cite{silva,efg-15}. 
Over the last few years it has become clear that the quench dynamics
in integrable systems is very unusual. While generic systems relax
locally to standard Gibbs distributions at effective temperatures
set by the initial energy density \cite{nonint,rdo-08,tvar},
integrable models retain a detailed memory of the initial state even
at infinite times \cite{kww-06,gg,rs-12,ce-13} by virtue of having
infinite sets of conserved charges.
It has been conjectured that in integrable systems expectation values
of local operators in the stationary state can be calculated using a 
generalized Gibbs ensemble (GGE)\cite{gg}, a statistical ensemble
formed from all local conserved charges (the role of locality of the
integrals of motion has been highlighted in Refs
\cite{calabrese_essler_fagotti_II,fe-13}).
%%%%%%%%%%%%%%%%%%
%%%%%%%%%%%%%%%%%%
%%%%%%%%%%%%%%%%%%
In order to understand and describe the behaviour of the quench
dynamics in interacting integrable models new approaches have been
developed \cite{ce-13,fm-10,gritsev,ia-12,cc-06,bse-14}.
Investigations of quench dynamics in such models have focussed on the
one-dimensional Bose gas (i.e. the Lieb-Liniger model)
\cite{cro,ck-12,de_nardis_wouters,m-13,ga-15}, the XXZ spin-chain
\cite{fe-13b,fcce-13,wouters_brockmann,pozsgay_werner_kormos,la-14}, the Richardson
model \cite{fcc-09} and sine- and sinh-Gordon field theories
\cite{stm-14,bse-14}.
For the Bose gas, the knowledge of the overlaps between the BEC state and the eigenstates of the system with arbitrary interaction 
\cite{de_nardis_wouters,cd-14} allowed for a characterization of the
stationary state \cite{de_nardis_wouters} which turned out to be
compatible with a properly constructed GGE \cite{ks-13}, while the
full time evolution of observables is analytically known only in the
limit of strong interaction \cite{kcc14,dc-14} (but it is possible to
exploit integrability to study the time evolution numerically
\cite{ce-13,de_nardis_wouters,dpc-15}). 
For the XXZ spin chain, the overlaps with the Bethe eigenstates are
known only for some classes of product states
\cite{p-13,brockmann_I,brockmann_II,brockmann_III,pc-14}. Again these
overlaps allowed for a description of the stationary
state in the gapped regime \cite{wouters_brockmann,pozsgay_werner_kormos}
which, surprisingly, turned out to disagree with the predictions of a
GGE formed by the infinite number of ultra-local conservation laws of
the XXZ chain \cite{fe-13b,fcce-13}. This disagreement motivated many
investigations aimed at understanding its origin, see
e.g. \cite{ga-14,ppsa-14,mpp-15, emp-15,imp-15,f-14b}, which
culminated in the recently proposed solution \cite{newGGE} based on taking
into account quasi-local conserved charges.

In spite of this intense research effort, relatively little is known
about the full time dependence of observables in interacting
integrable models in the thermodynamic limit. This is in stark contrast
to non-interacting theories (or models that map onto such)
\cite{cc-06,kcc14,dc-14,cazalilla,ic-09,ic-10,cic-12,bs-08,cef,scc-09,mc-12,fc-08,s-08,rsm-10,ir-10,gcg-11,se-12,csc-13,bkc-14}. 
In particular, there are no exact results for the behaviour after
quantum quenches in the gapless phase of the Heisenberg XXZ model
(with the exception of the XX chain, corresponding to non-interacting
fermions \cite{bar-09,f-14}). However, such quenches have been
analyzed in a number of numerical works
\cite{bar-09,dmcf-06,krs-12,cbp-13}. The critical phase of the
Heisenberg chain is of particular interest, because in equilibrium many
of its properties can be described by Luttinger liquid
theory \cite{LLtheory,rev-bos,lukyanov,hiki}. An obvious question is
then, to what extent, if any, Luttinger liquid theory can be used for
the analysis of quench dynamics. Given that the quantum quench deposits
an extensive amount of energy in the system, perturbations to the
Luttinger liquid description as well as cutoff effects are expected to
play a role. Before considering such issues one first ought to investigate
whether the simple results obtained for quantum quench in the
Luttinger model \cite{cazalilla} can effectively describe the
non-equilibrium dynamics of the spin chain, at least in some parameter
regimes and/or specific time windows. Indeed this question has been
addressed in Refs \cite{Uhrig09,krs-12} for the so-called Z-factor and in
Ref. \cite{cbp-13} for the stationary values of correlation functions
in momentum space.

Effects of perturbations away from the Luttinger model have been
analyzed by renormalization group methods \cite{mg-11,m-12}. This
gives some insight in how perturbations lead to eventual
thermalization on very long time scales. 
We mention that many other aspects of the quench dynamics of the
Luttinger model have been considered in the literature
\cite{p-06,lm-10,ps-11,rsm-12,dbz-12,HU13,phd-13,km-13,bd-13,ni-13,dbm-13,KKM-14,snad-14,svph-14,bd-15,dp-15}.
The aim of our work is to quantify in some detail to what extent ``simple''
Luttinger liquid theory can be used to describe the time evolution of
local observables after an interaction quench in the critical phase of
the spin-1/2 Heisenberg XXZ chain. To that end we compare predictions
of Luttinger liquid theory for spin-spin correlation functions to
numerical results. 

%%%%%%%%%%%%%%%%%%%%%%%%%%%%%%%%%%%%%%%%%%%%%%
\subsection{The model and the quench protocol}
%%%%%%%%%%%%%%%%%%%%%%%%%%%%%%%%%%%%%%%%%%%%%%
We consider is the one-dimensional spin-$1/2$ XXZ chain with Hamiltonian
\be\label{H_XXZ}
H(\Delta) = J \sum_{j=1}^{L} \left( S^{x}_{j}S^{x}_{j+1} + S^{y}_{j}S^{y}_{j+1} + \Delta\,S^{z}_{j}S^{z}_{j+1} \right ),
\ee
where $S^{\alpha}_{j}$ are spin operators at site $j$ of a one
dimensional chain and $\Delta$ parametrizes the exchange
anisotropy. Throughout this paper we will set $J=1$, which amounts to
measuring all energies in units of $J$. The equilibrium properties of
the system in the thermodynamic limit are well known:
for $|\Delta|<1$ the system is quantum critical, while for $|\Delta|>1$
the ground state is antiferromagnetically ordered and excitations
acquire an energy gap. 

In the following we want to study the out-of-equilibrium dynamics in
the gapless phase. Our protocol is as follows: we prepare the system
in the ground state $|\Psi_{0}\rangle$ of the XX model
($\Delta_{0}=0$) and at time $t=0$ suddenly quench the anisotropy
parameter $\Delta$ to a finite value in the interval $(-1,1]$ and then
time evolve unitarily with Hamiltonian $H(\Delta)$. This protocol is
often referred to as \emph{interaction quench}. We note that the
generalization of our work to an arbitrary quench
$\Delta_0\rightarrow\Delta$ of the anisotropy parameter within the
gapless phase is straightforward. In the following we present a
systematic numerical study of the quench dynamics and compare the
numerical results to analytic expressions for correlation functions
obtained via Luttinger liquid theory (LL). 

%%%%%%%%%%%%%%%%%%%%%%%%%%%%%%%%%%%%%%%%%%%%%
\subsection{Organization of the manuscript} 
%%%%%%%%%%%%%%%%%%%%%%%%%%%%%%%%%%%%%%%%%%%%%
In Sec. \ref{sec:bos} we review the bosonization of the XXZ
Hamiltonian and summarize how to use Luttinger liquid theory to obtain
results for long-distance behaviour of spin-spin correlation functions
after an interaction quench.
In Sec. \ref{sec:num} we present the results of numerical computations
of the same quantities by means the time-evolving block decimation algorithm.  
The numerical results are carefully compared with the bosonization
predictions. Finally, in Sec. \ref{concl} we summarize our results.

%%%%%%%%%%%%%%%%%%%%%%%%%%%%%%%%%%%%%%%%%%%%%%%%%%%%%%%%%%%%%%%%%%
\section{Bosonization of the XXZ spin chain}
\label{sec:bos}
%%%%%%%%%%%%%%%%%%%%%%%%%%%%%%%%%%%%%%%%%%%%%%%%%%%%%%%%%%%%%%%%%%

In this section we summarize some key results on the low energy field
theory description of the spin-1/2 XXZ chain in zero magnetic field,
details can be found in e.g. \cite{rev-bos}. At low energies the
spin-1/2 Heisenberg chain can be described as a free compact boson
perturbed by irrelevant operators
\be
{\cal H}(\Delta) =\frac{v}{2} \int dx \left[K (\partial_x\theta)^2
+\frac{1}K (\partial_x\phi)^2\right]+{\cal H}_{\rm irr}  .
\ee
Here $\phi=\varphi+\bar\varphi$ is a canonical Bose field,
$\theta=\varphi-\bar\varphi$ the dual field ($\varphi$ and
$\bar\varphi$ are chiral components), $v$ is the spin velocity, $K$ is
the so-called Luttinger parameter, and ${\cal H}_{\rm irr}$ denotes an infinite
tower of perturbing operators that are irrelevant in the
renormalization group sense. In zero field the values of $K$ and
$v=\vt a_0$ (we recall that we have set $J=1$) are known exactly from
the Bethe ansatz solution of the XXZ chain \cite{book-ba} 
\be
\tilde{v}=  \frac{\pi}{2}\frac{\sqrt{1-\Delta^2}}{\arccos \Delta},\qquad
K  = \frac{\pi}{2}\frac{1}{\pi-\arccos \Delta },
\label{KvBA}
\ee
where $a_0$ is the lattice spacing. In the XX limit $\Delta=0$ we have
$\vt=1$ and $K=1$. The bosonized expressions for the lattice spin
operators are \cite{rev-bos,lukyanov,hiki}
\bea
S^z_j&\simeq& m-\frac{a_0}{\sqrt{\pi }}\partial_x\phi(x)
+(-1)^j a_1 \,\sin (\sqrt{4 \pi } {\phi(x)})+\ldots,  \label{Szbos} \\
S^{x}_j&\simeq& b_0(-1)^j\cos\big(\sqrt{\pi} \,\theta(x)\big)
+ib_1\sin\big(\sqrt{\pi} \,\theta(x)\big)
 \sin\big({\sqrt{4 \pi }\phi(x)}\big)+\ldots \, \label{S-bos},
\eea
where $x=ja_0$ and $m= \langle S^z_j\rangle$.
Using these expressions one arrives to the equilibrium two-point functions 
\bea
\langle S^{x}_{j} S^{x}_{j+\ell} \rangle&=& (-1)^\ell \frac{A_0^x}{\ell^{1/(2K)}} - A_1^x \frac{1}{\ell^{2K+1/(2K)}}, \\
\langle S^{z}_{j} S^{z}_{j+\ell} \rangle&=& m^2 -\frac{2K}{\pi^2} \frac1{\ell^2} + A_1^z \frac{(-1)^\ell}{\ell^{2K}}.
\eea

In the absence of a magnetic field ($m=0$) the various non-universal constants are
known. In two-point functions the relevant amplitudes are
\be
A^x_0=\left|\frac{b_0^2}{2}\right|\ ,\quad
A^x_1=\left|\frac{b_1^2}{4}\right|\ ,\quad
A^z_1=\left|\frac{a_1^2}{2}\right|\ .
\ee
Their values are given by \cite{lukyanov} 
\bea
\fl A^x_0&=&
\frac{1}{8(1-\eta)^2}
\left[\frac{\Gamma(\frac{\eta}{2(1-\eta)})}
            {2\sqrt{\pi}\,\Gamma(\frac{1}{2(1-\eta)})}
\right]^\eta
%\nonumber\\
%&&\times 
\exp\left[
-\int^\infty_0\frac{dt}{t}
  \left(\frac{\sinh(\eta t)}{\sinh(t)\cosh[(1-\eta)t]}
        -\eta e^{-2t}\right)\right],
\label{eq:A0xLuky}
\eea
\bea
\fl
A^x_1 =
\frac{1}{2\eta(1-\eta)}
\left[\frac{\Gamma(\frac{\eta}{2(1-\eta)})}
            {2\sqrt{\pi}\,\Gamma(\frac{1}{2(1-\eta)})}
\right]^{\eta+\frac{1}{\eta}}\nonumber\\
\fl\times\exp\left[
-\int^\infty_0\frac{dt}{t}
  \left(\frac{\cosh(2\eta t)e^{-2t}-1}{2\sinh(\eta t)\sinh(t)\cosh[(1-\eta)t]}
     +\frac{1}{\sinh(\eta t)}-\frac{\eta^2+1}{\eta} e^{-2t}\right)\right],
\label{eq:A1xLuky}
\eea
\bea\fl
A^z_1 &=&
\frac{2}{\pi^2}
\left[\frac{\Gamma(\frac{\eta}{2(1-\eta)})}
            {2\sqrt{\pi}\,\Gamma(\frac{1}{2(1-\eta)})}
\right]^{\frac{1}{\eta}}
%\nonumber\\
%&&\times
\exp\left[
  \int^\infty_0\frac{dt}{t}
  \left(\frac{\sinh[(2\eta-1) t]}{\sinh(\eta t)\cosh[(1-\eta)t]}
        -\frac{2\eta -1}{\eta} e^{-2t}\right)\right].
\label{eq:A1zLuky}
\eea
Here the parameter $\eta$ is related to the anisotropy $\Delta$ by
\be
\eta=1-\frac{1}{\pi}{\rm arccos}(\Delta)=\frac1{2K}.
\ee
%%%%%%%%%%%%%%%%%%%%%%%%%%%%%%%%%%%%%%%%%%%%%%%%%%%%%%%%%%%%%%%%%%
\subsection{Interaction quench}
\label{ssec:Iquench}
%%%%%%%%%%%%%%%%%%%%%%%%%%%%%%%%%%%%%%%%%%%%%%%%%%%%%%%%%%%%%%%%%%
Our quench protocol is to prepare our spin chain in the ground state
of the XX-chain, i.e. the Hamiltonian (\ref{H_XXZ}) with $\Delta=0$,
and then time evolve the system with $H(\Delta)$ at times $t>0$. This
corresponds to a sudden quench of the interaction strength from
$\Delta=0$ to $\Delta\in (-1,1]$ at time $t=0$. Assuming that a projection
to the low-energy description in terms of a Luttinger liquid is
possible, this corresponds to preparing our system in the ground state
of the free boson theory ${\cal H}(\Delta=0)$, and then time evolve it
with ${\cal H}(\Delta)$. The observables of interest are the
low-energy projections (\ref{Szbos}), (\ref{S-bos}) of the spin
operators. Two point functions of these operators have been calculated
for interaction quenches in the Luttinger model in
Refs.~\cite{cazalilla,ic-09}, and can be used in the case of interest
here. In order to make our discussion self-contained we summarize the
main points of the necessary calculations in~\ref{app:LL}.
The final results of these calculations are 
\be\label{Eq_LM_SxSx}
\langle S^{x}_{j}(t) S^{x}_{j+\ell}(t) \rangle 
\simeq (-1)^{\ell} \frac{A^{x}}{\sqrt{\ell}} \left|  \frac{1}{(2\vt
  t)^2} \frac{\ell^2-(2\vt t)^2}{\ell^2} \right|^{(1/K^2-1)/8}+\cdots,
\ee
and
\bea
\fl \langle S^{z}_{j}(t) S^{z}_{j+\ell}(t) \rangle  & \simeq &   
B_z\left\{-\frac{1-K^2}{8\pi^2}  \left[\frac{1}{(\ell+2\vt t)^2} +
  \frac{1}{(\ell-2\vt t)^2}\right] -
\frac{1+K^2}{4\pi^2}\frac{1}{\ell^2} \right\}\nn
&&+A^{z}
\frac{(-1)^{\ell}}{\ell^2} \left|  \frac{1}{(2\vt t)^2}
\frac{\ell^2-(2\vt t)^2}{\ell^2} \right|^{(K^2-1)/2} +\cdots.
\label{Eq_LM_SzSz}
\eea
In the correlator $\langle S^{x}_{j}(t) S^{x}_{j+\ell}(t) \rangle$  we
retain only the staggered term because the other contributions
turn out to be small, while for $\langle S^{z}_{j}(t)
S^{z}_{j+\ell}(t) \rangle$ the smooth and staggered terms
turn out to be comparable in magnitude. As first pointed out
in Ref.~\cite{cazalilla} for the Luttinger model, the power-law decays
(\ref{Eq_LM_SxSx}), (\ref{Eq_LM_SzSz}) are very different from their
equilibrium analogs. The amplitudes $A^{x/z}$ and $B^{z}$ are
non-universal and will be fixed below by fitting (\ref{Eq_LM_SxSx}),
(\ref{Eq_LM_SzSz}) to numerical results. They are related to the
equilibrium amplitudes $A_0^{x}$ and $A^1_{z}$ via unknown relations
involving the cutoffs of pre- and post-quench Hamiltonians. If the
quench does not require a cutoff adjustment one would expect that
\be
A^x\sim A^x_0, \qquad  A^z\sim A^z_1, \qquad B^z\sim1\,.
\ee
We will see that these relations turn out to be approximately
fulfilled, with deviations of the order of a few percent.  

As (\ref{Eq_LM_SxSx}) and (\ref{Eq_LM_SzSz}) are derived in the
framework of a field theory approximation they are applicable to the
description of lattice correlators only as long  as $\ell,
vt,|\ell-2vt| \gg a_0=1$.
In Table \ref{tab1} we summarize the numerical values of the
parameters for the post-quench interaction strengths that we will
consider in the following,  i.e. $\Delta = -0.5, \, -0.2, \, 0.2, \,
0.5$.

%%%%%%%%%%%%%%%%% TABLE LUTTINGER LIQUID PARAMETERS %%%%%%%%%%%%%
\begin{table}[t!]
\center
\begin{tabular}{|c | c c c c |}
\hline
$\Delta$ & $K$ & $v$ & $\alpha$ & $\beta$\\
\hline
$-0.5$ & $3/2$ & $3\sqrt{3}/8$ & $-5/36$ & $5/4$\\

$-0.2$ & $1.14704$ & $0.868468$ & $-0.0599861$ & $0.315694$ \\

$0$ & $1$ & $1 $ & $0$ & $0$\\

$0.2$ & $0.886377$ & $1.12386$ & $0.0682023$ & $-0.214336$ \\

$0.5$ & $3/4$ & $3\sqrt{3}/4$ & $7/36$ & $-7/16$\\

%$0.7$ & $0.669508$ & $1.41033$ & $-0.412063$  & $0.307735$\\
\hline
\end{tabular} 
\caption{Luttinger liquid parameters for the values of $\Delta$ considered here.}\label{tab1}
\end{table}
%%%%%%%%%%%%%%%%%%%%%%%%%%%%%%%%%%%%%%%%%%%%%%%%%%%%%%

%They also present power-law singularity on the light cone $\ell=2vt$ which are non-physical.
%In the spin chain (and in any other lattice model), these singularity are smoothed on the scale of the lattice spacing $a$
%(and so we also require $\gg a_0$).
%One could easily cure these singularities by introducing a further fitting parameter (at the scale of the lattice spacing).
%This issue will be discussed in the analysis of the numerical data. 

%%%%%%%%%%%%%%%%%%%%%%%%%%%%%%%%%%%%%%%%%%%%%%%%%%%%%%%%%%%%%%%%%%
\section{Quantum quenches in the spin-$1/2$ XXZ chain: from free
fermions to the quantum critical phase} 
\label{sec:num}
%%%%%%%%%%%%%%%%%%%%%%%%%%%%%%%%%%%%%%%%%%%%%%%%%%%%%%%%%%%%%%%%%%

We now turn to the central part of this paper, a detailed numerical
study of the quench dynamics of the XXZ spin-chain from $\Delta_0=0$
to a final $-1<\Delta\leq 1$ in the gapless phase.
To this end we employ the infinite time-evolving block decimation
(iTEBD) algorithm \cite{iTEBD} to study the dynamics induced by the
post quench Hamiltonian (\ref{H_XXZ}).  
The algorithm is based on a matrix product state (MPS) description
of one-dimensional lattice models and works directly in the
thermodynamic limit. Compared to other algorithms like time-dependent
density matrix renormalization group \cite{tDMRG}, the iTEBD has the
great advantage of not introducing systematic errors due to finite size
effects, since it works directly in the thermodynamic limit. 
This is made possible by two main features: (i) invariance under
translation of the Hamiltonian; (ii) the possibility to parallelize
the local updates of the time-evolving block decimation procedure. 
 
%%%%%%%%%%%%%%%
\subsection{Method}
%%%%%%%%%%%%%%%
The iTEBD algorithm we are using is composed of two different parts: 
the first obtains an accurate MPS description of the initial state (in
our case the ground state of the XX Hamiltonian) and the second deals with
the time evolution.  

The ground state  of the XX Hamiltonian $H_{XX}\equiv H(\Delta=0)$ is
found using the iTEBD algorithm in imaginary time. The infinite chain
is prepared in a given, simple state $|\Psi_s\rangle$, which we choose
as $|\Psi_s\rangle=\bigotimes_{j\in\mathbb{Z}} (|\uparrow\rangle_{j} +
|\downarrow\rangle_{j})/\sqrt{2}$. This state has a trivial MPS  
representation of bond dimension $\chi_{0}=1$. 
We then evolve $|\Psi_s\rangle$ in imaginary time with $\exp (-\tau
H_{XX})$, implemented using the second order Suzuki-Trotter
decomposition with imaginary time-step $\tau = 0.01$. We verified
that further decreasing $\tau$ does not affect the final result within
our numerical accuracy. As is well known, since the imaginary-time
evolution operator is not unitary, the MPS loses its canonical form
(i.e. its normalisation is not constant). Hence this form must be
restored while time elapses and in particular before taking the
expectation value of any operator.
We control the convergence of the imaginary time algorithm by keeping
track of the energy density $E_{0}$ and waiting for it to become
stationary (with an accuracy of $10^{-16}$). We repeat this
procedure for several values of the bond dimension up to $\chi_0 =
128$. A check of this procedure is provided by our best estimate
of the ground-state energy density, which is $E_{0} = -0.31830981$.
This differs from the exact value $E_{XX} =  -1/\pi$ by $7\cdot
10^{-8}$. This is quite satisfactory as the XX model is gapless and
exhibits long range correlations. An exact MPS representation of the
ground state in this case requires an infinite bond dimension (in
other words, the singular values involved in the MPS show an algebraic
decay). For our purposes the description of the initial state with
$\chi_0=128$ is sufficiently accurate. Indeed, comparing the
longitudinal and transverse spin-spin correlation functions with the
exact analytical results up to distances of $\ell = 50$, we find an
extremely good agreement, with relative errors smaller than  
$3\%$ and $1\%$ for respectively longitudinal and transverse
correlators.

Using this MPS as our initial state, we can address the real time
evolution. We again use a second order Suzuki-Trotter decomposition of
the evolution operator $\exp(-i dt H)$ with time step $dt = 0.05$  
(we verified that the data are not affected by the time discretisation).
We adapt the number of states used to describe the reduced Hilbert
space by retaining, at each time step, all Schmidt vectors
corresponding to singular values larger than $\lambda_{min}=10^{-12}$,
up to a maximum value $\chi_{\rm MAX}=1024$. The latter is actually
reached fairly quickly, reflecting the fast growth of the entanglement
entropy under time evolution. Indeed, it is well known that the
computational complexity of the time evolution of a quantum system  
using algorithms based on MPS descriptions is essentially set by the
growth of the bipartite entanglement. As the entanglement increases
with time, we have to enlarge the dimension $\chi$ of the reduced
Hilbert space in order to optimally control the truncation error. 
In spite of our refined adaptive choice of $\chi$, the truncation
procedure is still the main source of error in our algorithm.  
For this reason, we are able to reach a maximum time $T = 20$ without
significant truncation error (the Schmidt error coming from the
iterative truncation of the Hilbert space remains always smaller than
$2\cdot 10^{-3}$). This is  also reflected on the behaviour of the
entanglement entropy of half system, which grows linearly for all
explored times as it should after a global quantum quench
\cite{cc-05-quench,fc-08}.

%%%%%%%%%%%%%%%%%%%%%%%%%%%%%%%%%%%%%%%%%%%%%%%%%%%%%%%%%
\section{Transverse correlation functions $\langle S^x_i S^x_j\rangle$}
%%%%%%%%%%%%%%%%%%%%%%%%%%%%%%%%%%%%%%%%%%%%%%%%%%%%%%%%%
We now turn to the two point correlation function of the x-component
of spin. We find that $\langle S^{x}_{j}S^{x}_{j+\ell}\rangle$
displays a strongly alternating behaviour in space, i.e. it is
dominated by a contribution of the form $(-1)^\ell f(\ell)$, where
$f(\ell)$ is a smooth function. Given that this structure is correctly
predicted by the Luttinger model, cf. (\ref{Eq_LM_SxSx}), we will
only analyze the absolute value $|\langle S^{x}_{i}S^{x}_{j}\rangle|$
(as this makes plotting the correlation function easier).

%%%%%%%%%%%% FIGURE SxSx   vs t   Delta > 0 %%%%%%%%%%%%%
\begin{figure}[t]
\center
\includegraphics[width=0.45\textwidth]{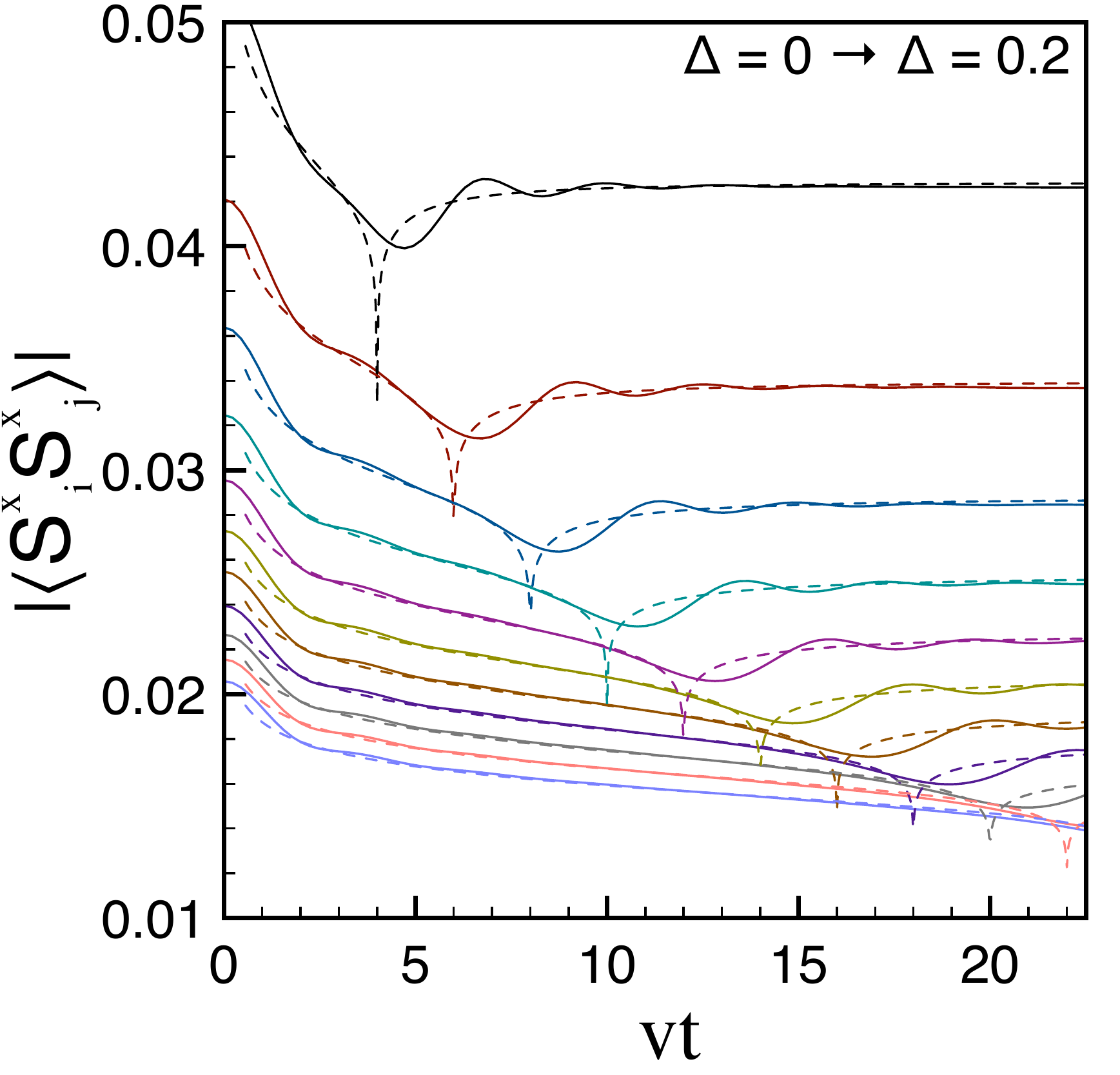}
\includegraphics[width=0.45\textwidth]{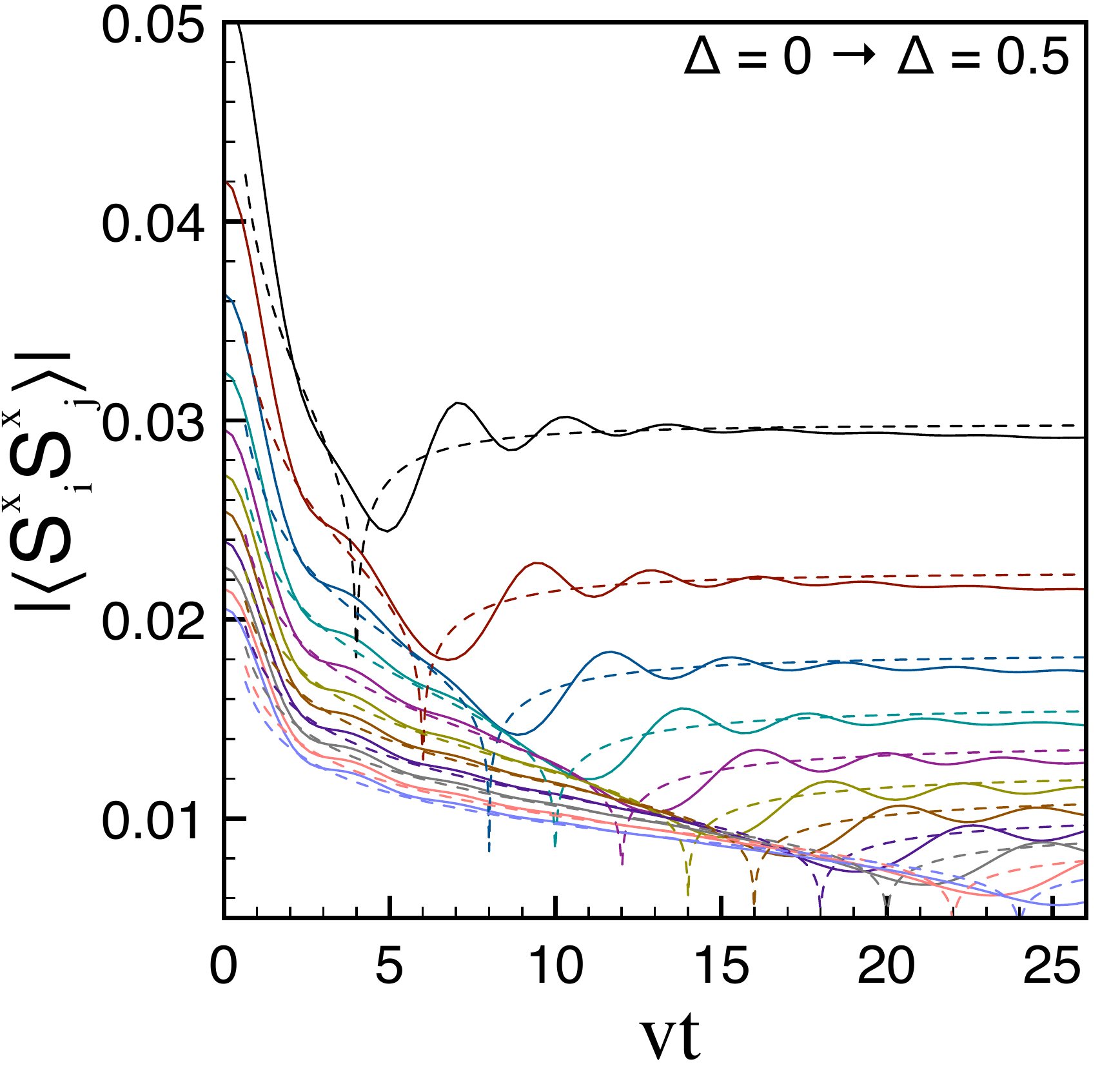}\\
\includegraphics[width=0.45\textwidth]{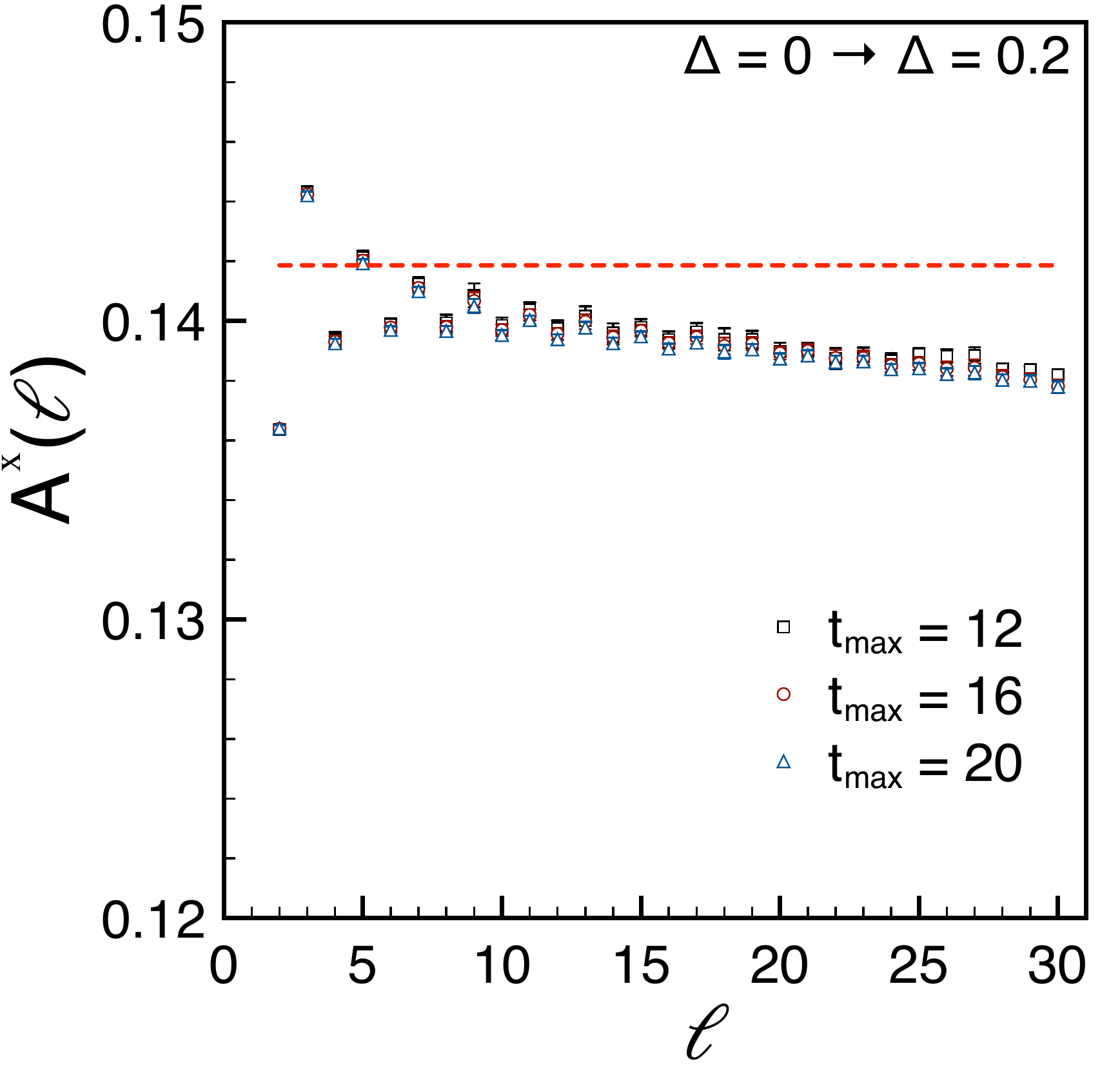}
\includegraphics[width=0.45\textwidth]{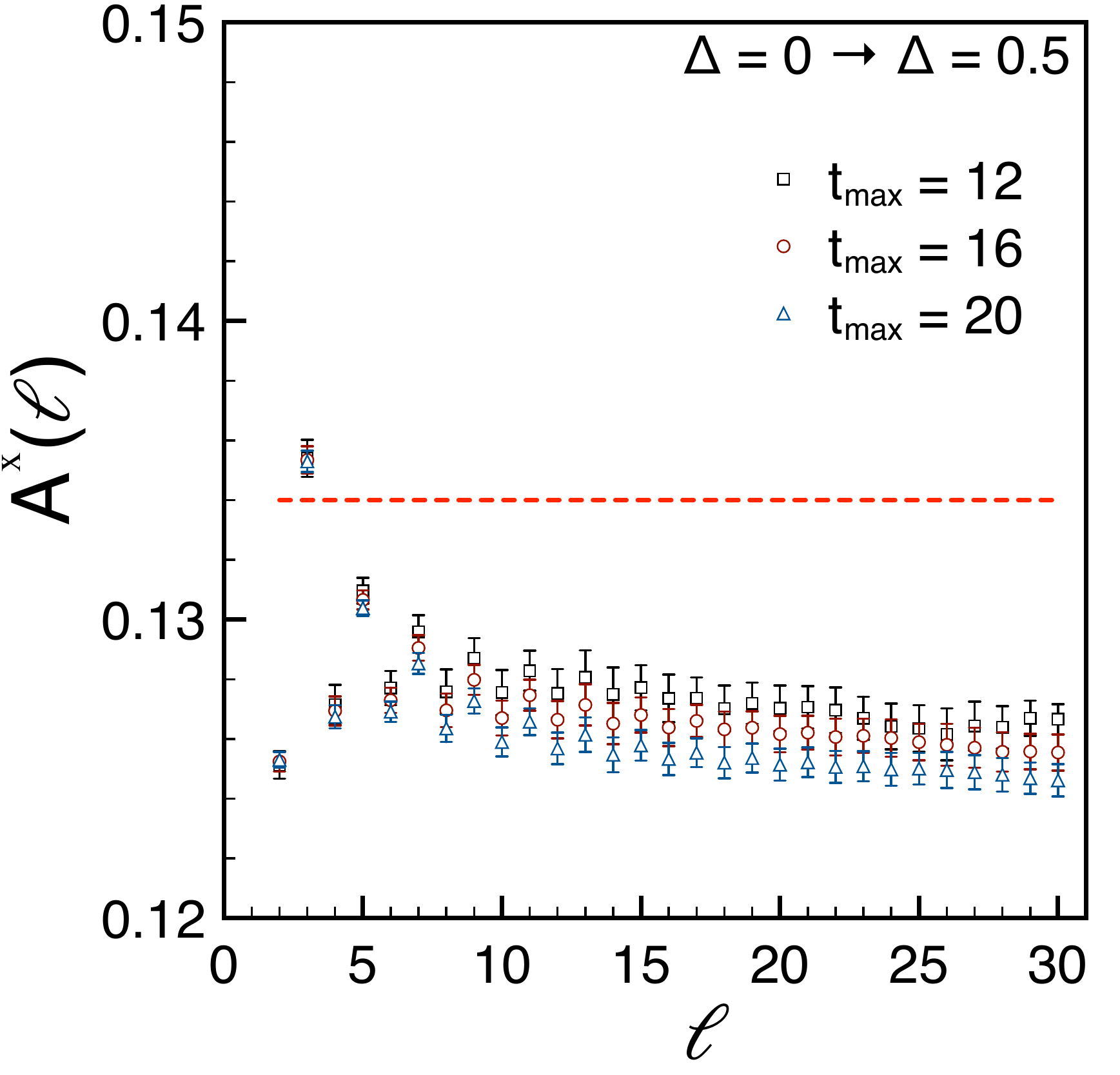}\\
\caption{\label{figSxSx_vs_t_1}
(Top) $|\langle S^{x}_{i}S^{x}_{j}\rangle|$ vs time, for different distances
$\ell =j-i= 8,\, 12,\, 16,\, \dots ,\, 48$  (from top to bottom)  and
interaction strengths $\Delta>0$. Full lines are the numerical data,
dashed lines represent the best fits to (\ref{Eq_LM_SxSx}) on the
interval $0\leq t<t_{max} =  20$. (Bottom) The pre-factor $A^{x}$
obtained from the best fit for different values of $t_{max}$. The
dashed line is the equilibrium value of the amplitude $A_0^x$ in
Eq. (\ref{eq:A0xLuky}).} 
\end{figure}
%%%%%%%%%%%%%%%%%%%%%%%%%%%%%%%%%%%%%%%%%%%%%%%

%%%%%%%%%%%% FIGURE SxSx   vs t   Delta < 0 %%%%%%%%%%%%%
\begin{figure}[t]
%\begin{figure}[t!]
\center
\includegraphics[width=0.45\textwidth]{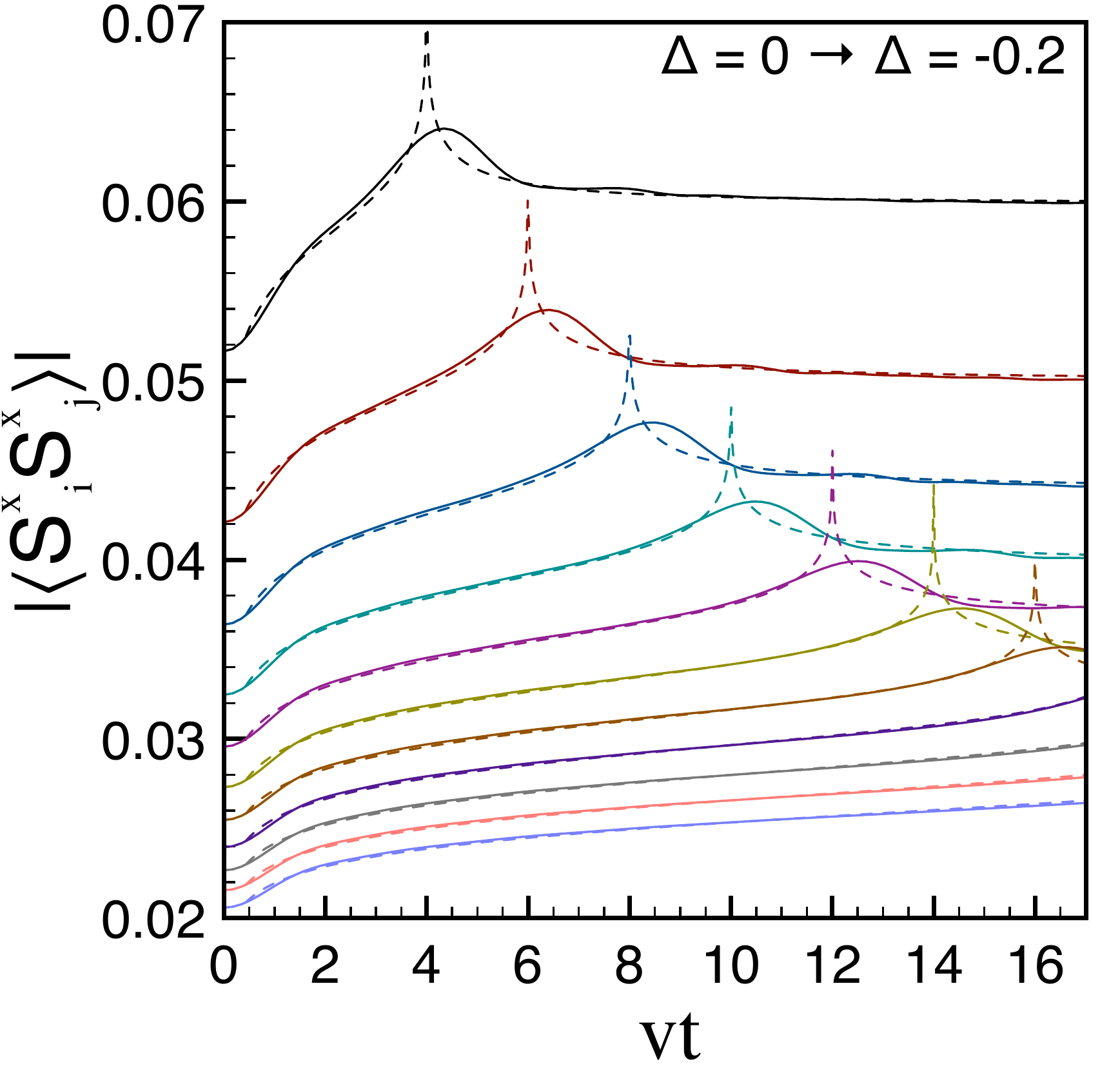}
\includegraphics[width=0.45\textwidth]{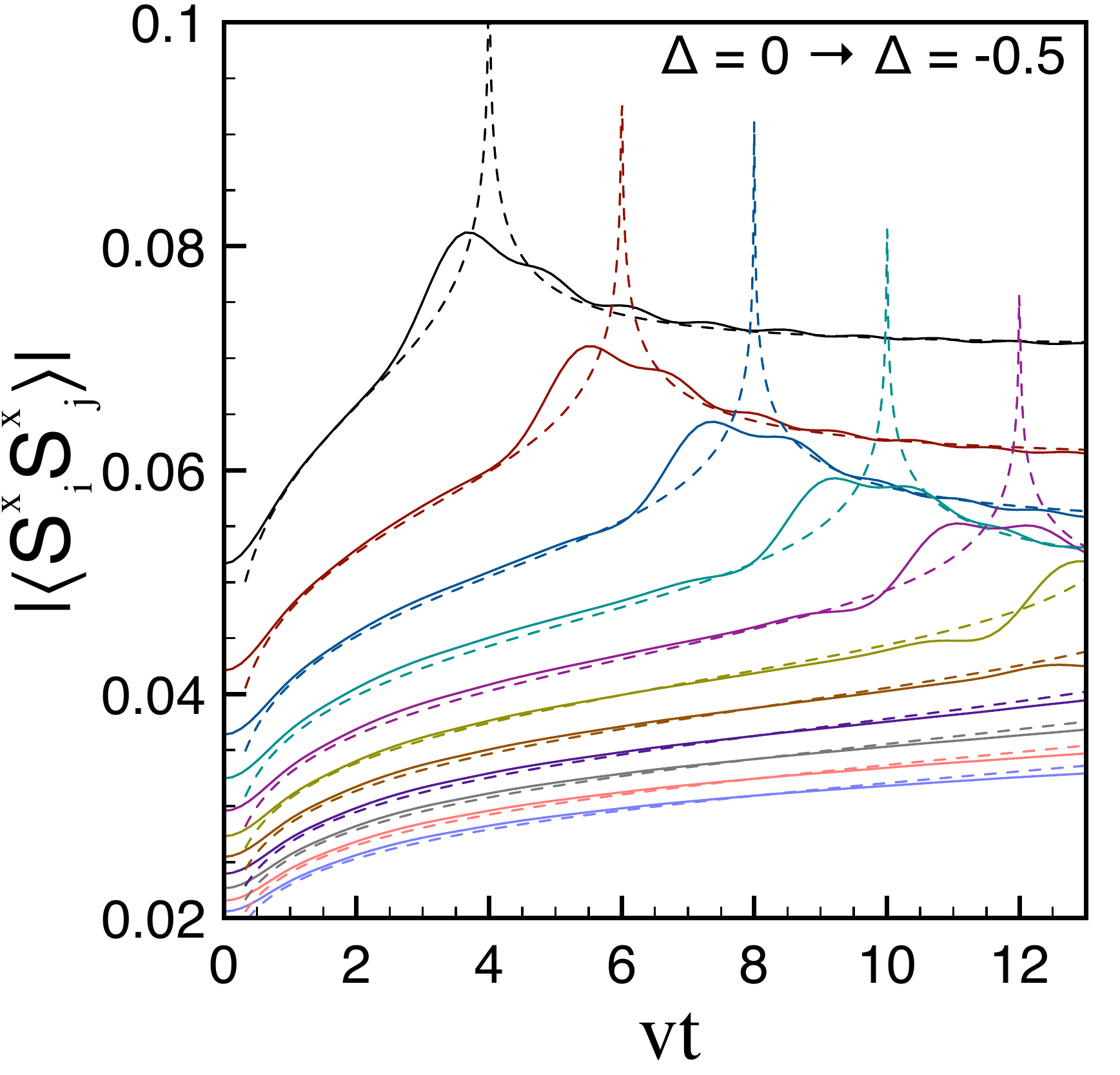}\\
\includegraphics[width=0.45\textwidth]{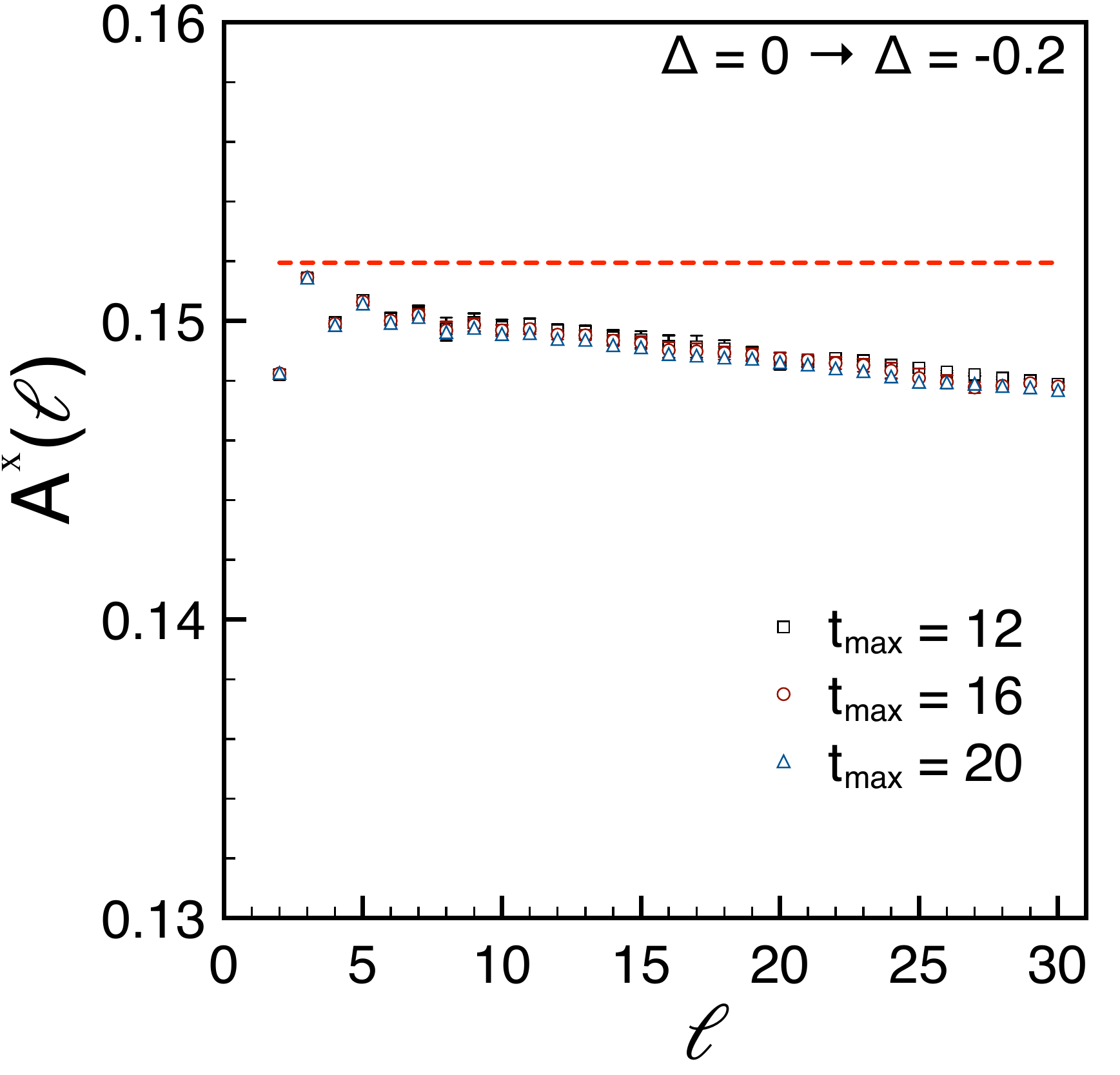}
\includegraphics[width=0.45\textwidth]{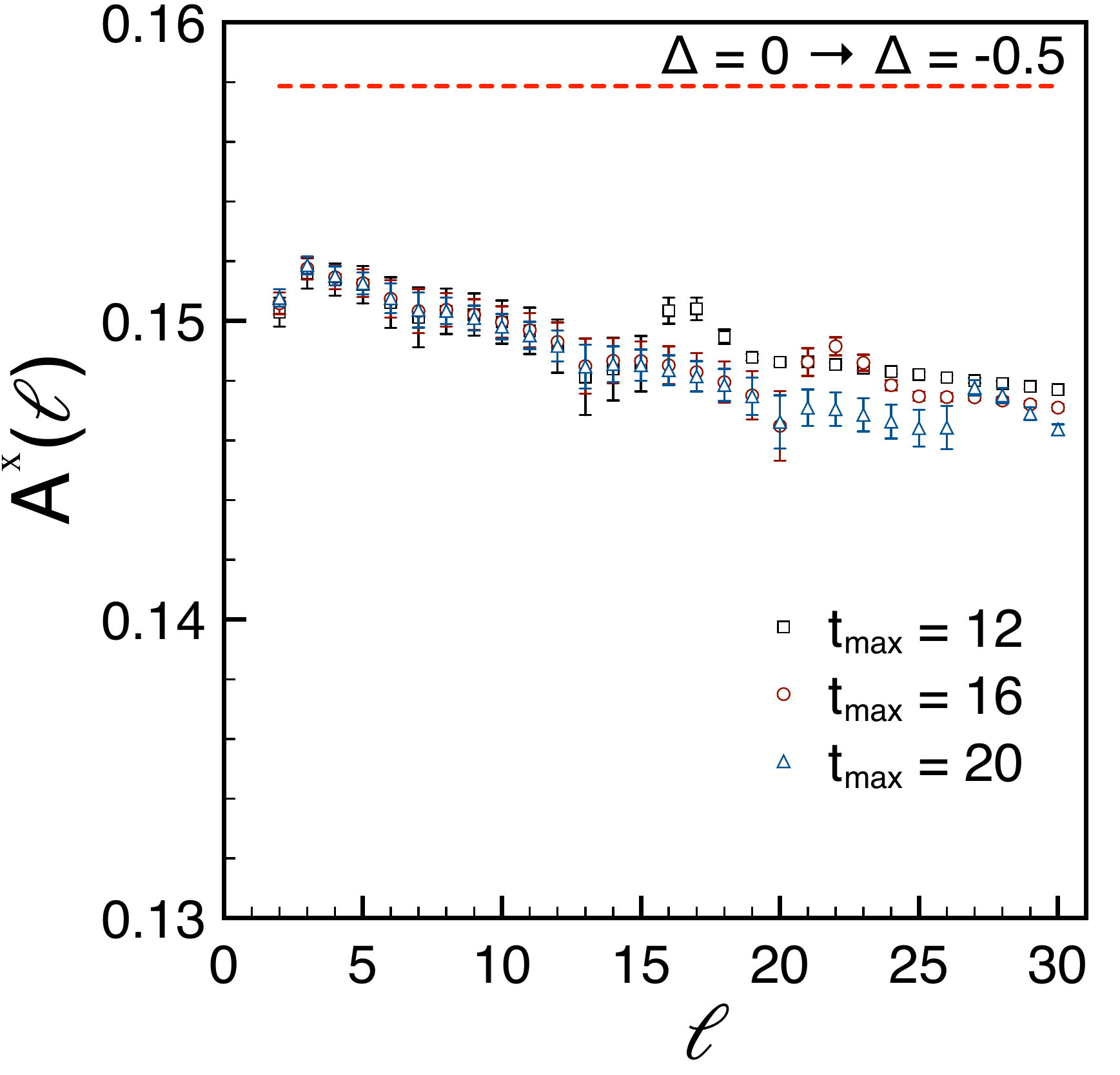}\\
\caption{\label{figSxSx_vs_t_2}
The same as in Figure \ref{figSxSx_vs_t_1} for $\Delta <0$.} 
\end{figure}
%%%%%%%%%%%%%%%%%%%%%%%%%%%%%%%%%%%%%%%%%%%%%%%

%%%%%%%%%%%% FIGURE  SxSx   vs x %%%%%%%%%%%%%
\begin{figure}[t!]
\center\includegraphics[width=0.75\textwidth]{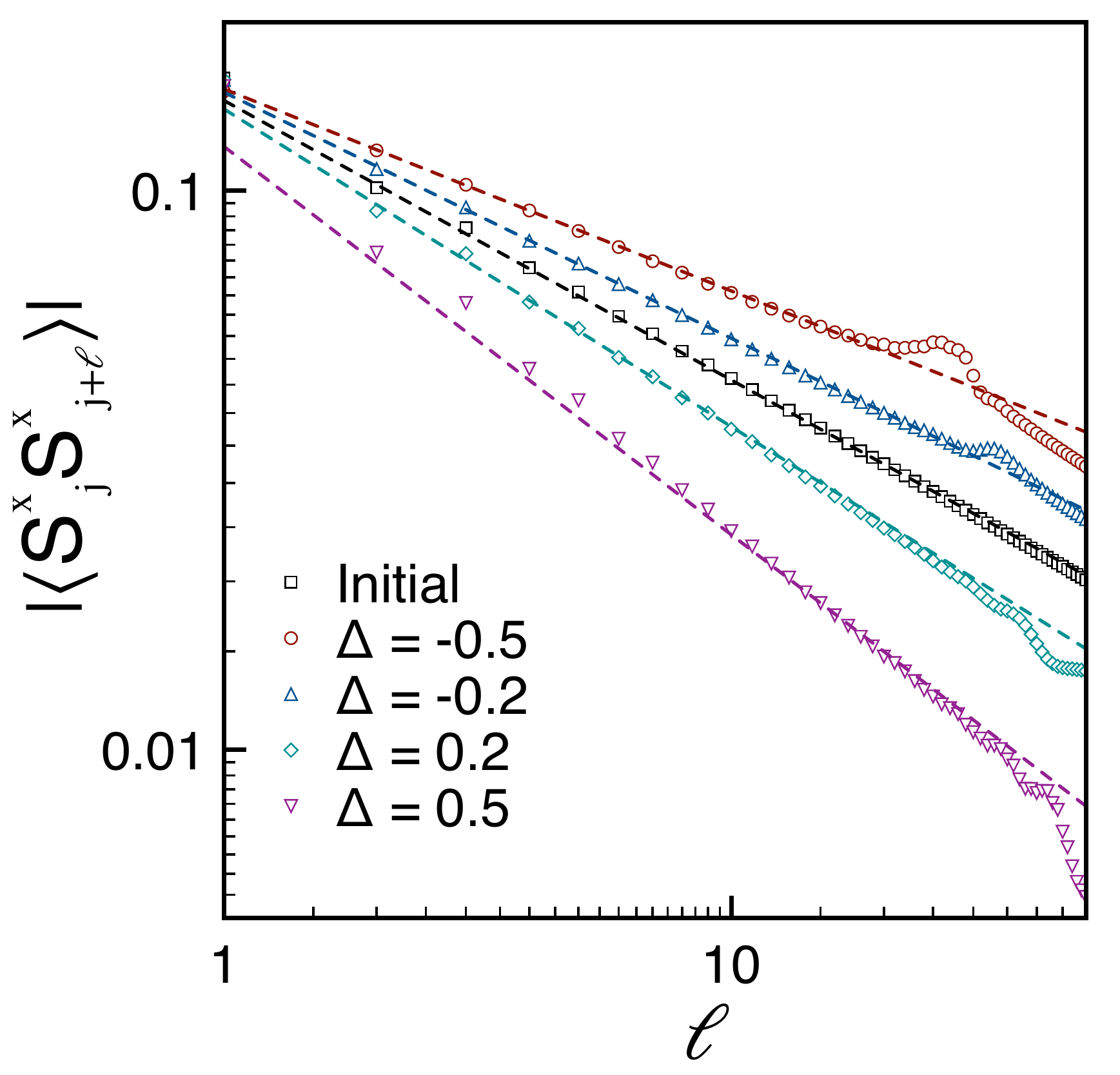}
\caption{\label{figSxSx_vs_x} Correlation function $|\langle
S^{x}_{j}S^{x}_{j+\ell}\rangle|$ at time $t = 20$, plotted as a
function of the lattice distance $\ell$, for different interaction
strengths $\Delta$ (different symbols). For comparison we also report
the initial correlators (black squares). The dashed lines represent
the asymptotic LL predictions $\sim \ell^{-\alpha-1/2}$.}  
\end{figure}
%%%%%%%%%%%%%%%%%%%%%%%%%%%%%%%%%%%%%%%%%%

Our numerical results for $|\langle S^{x}_{j}S^{x}_{j+\ell}\rangle|$
are shown in Figs \ref{figSxSx_vs_t_1} and \ref{figSxSx_vs_t_2} as
functions of time for different, fixed distances $\ell$, and in Fig.~
\ref{figSxSx_vs_x} as functions of $\ell$ for the largest time $t=20$.  
The main qualitative features of this correlator are as follows.
(i) Apart from a brief, non universal transient, $|\langle
S^{x}_{j}S^{x}_{j+\ell}|\rangle$ is surprisingly well described by the
Luttinger liquid prediction (\ref{Eq_LM_SxSx}).

(ii) For all quenches the correlator exhibits a clear light-cone
effect. At ``short'' times $1\ll 2vt \ll \ell$, the correlations decay
in a power law fashion $\sim t^{-\alpha}$ with exponent $\alpha =
(1/K^2-1)/4$, in agreement with (\ref{Eq_LM_SxSx}). We note that the
exponent $\alpha$ is related to properties of the $Z$-factor studied
in Ref. \cite{krs-12}. The exponent $\alpha$ is positive (negative)
for positive (negative) interactions $\Delta$ and indeed a
qualitatively  different behaviour is evident just by looking at Figs.
\ref{figSxSx_vs_t_1} and \ref{figSxSx_vs_t_2}. 
For late times $2vt > \ell$ the correlation function exhibits a
power-law decay in time towards its asymptotic stationary value, which
scales with the distance as $\ell^{-\alpha-1/2}$ (cf. Eq. (\ref{Eq_LM_SxSx})).
The algebraic decay as a function of distance at late times is
shown in Fig. \ref{figSxSx_vs_x} for several quenches. Another feature
visible in Fig.~\ref{figSxSx_vs_x} is the interaction
strength dependence of the velocity, at which correlations spread.
This is reflected in the presence of ``bumps'' in the spatial dependence
of correlation functions in Fig. \ref{figSxSx_vs_x} at the positions
of the light cones $\ell^{*} \simeq 2v t$. Inspection of the numerical
results for different values of $\Delta$ shows that $\ell^*$, and hence
$v$, depends on $\Delta$. At distances $\ell>\ell^{*}$, the
correlators still display the power-law decay $\ell^{-1/2}$ of the
initial state. In particular for the quenches to $\Delta = -0.5$ and
$\Delta= -0.2$ (where the propagation velocity is lower) these
regions are visible in Fig. \ref{figSxSx_vs_x}. 

%%%%%%%%%%%%%%%%% TABLE COMPARISON FIT PARAMETERS %%%%%%%%%%%%%
\begin{table}[b]
\center
\begin{tabular}{|c | c c c |}
\hline
$\Delta$ & $\delta A^{x}$ & $\delta A^{z}$ & $\delta B^{z}$ \\
\hline
$-0.5$ & $6\%$ & $60\%$ & $5\%$ \\

$-0.2$ & $2\%$ & $10\%$ & $5\%$  \\

$0.2$ & $2\%$ & $10\%$ & $20\%$  \\

$0.5$ & $6\%$ & $6\%$ & $15\%$\\
\hline
\end{tabular} 
\caption{Approximate relative deviations of the fitting parameters from the {\it equilibrium} Luttinger   
values in Eqs. (\ref{eq:A0xLuky}) and  (\ref{eq:A1zLuky}). 
The large relative error in $A^{z}(\ell)$ for $\Delta = -0.5$ is due to the
fact that equilibrium value is very close to zero (see Fig. \ref{figSzSz_vs_t_short2}). 
 }\label{tab2}
\end{table}
%%%%%%%%%%%%%%%%%%%%%%%%%%%%%%%%%%%%%%%%%%%%%%%%%%%%%% 

Having established the main qualitative features of the transverse
correlator, we now turn to a quantitative analysis of the crossover
between the two regimes described by Eq. (\ref{Eq_LM_SxSx}). To that end
we perform linear fits of our numerical results to (\ref{Eq_LM_SxSx}) as a
function of time for fixed distances $\ell$. The parameters $K$ and
$v$ are fixed by the quench parameter $\Delta$ and reported in Table
\ref{tab1}. We repeat the fit procedure for several time-windows
$[1,t_{max}]$ with $t_{max}=12, 16, 20$. The fit parameter $A^{x}$ is
found to depend only weakly on the choice of $t_{max}$, see
Figs~\ref{figSxSx_vs_t_1} and \ref{figSxSx_vs_t_2}). We find that
the values of $A^x$ determined in this way are rather close to their
``equilibrium'' values (\ref{eq:A0xLuky}), with deviations of the
order of a few percent. In Table \ref{tab2} we report estimates of the
relative difference $\delta A^x\equiv \overline{{(A^x(\ell)-A^x_0)/A_0^x}}$ for all
considered quenches, where we perform an average over the values of  $\ell$
which we consider large enough. 
On the other hand, fixing $A^x$ to its equilibrium value results in significantly poorer
agreement between our numerical results and the Luttinger liquid
prediction (\ref{Eq_LM_SxSx}).

%%%%%%%%%%%%%%%%%%%%%%%%%%%%%%%
\subsection{Scaling behaviour}
%%%%%%%%%%%%%%%%%%%%%%%%%%%%%%%

%%%%%%%%%%%% FIGURE SxSx   vs 2vt/x (space/time scaling) %%%%%%%%%%%%%
\begin{figure}[t]
\center
\includegraphics[width=0.45\textwidth]{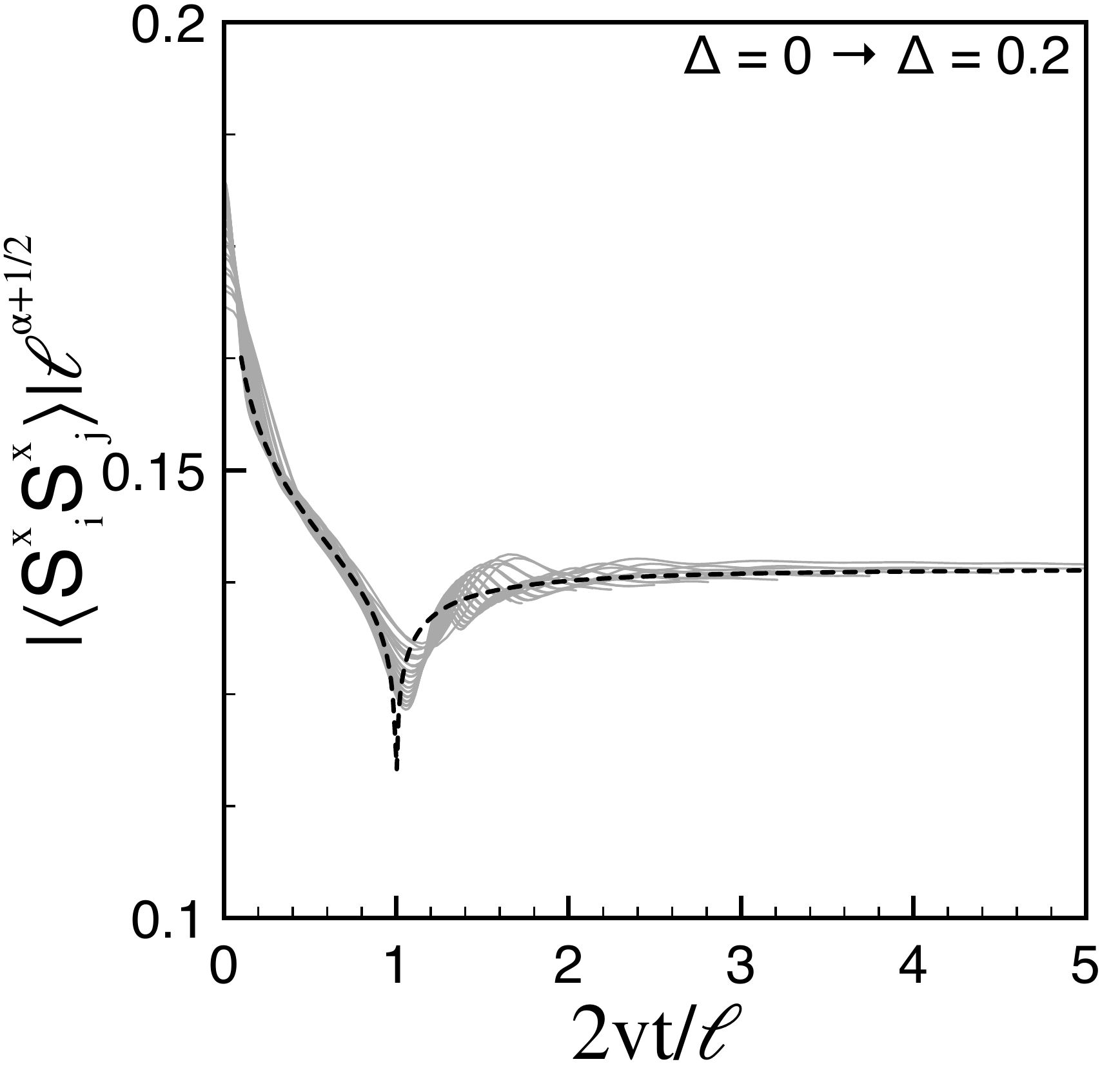}\includegraphics[width=0.45\textwidth]{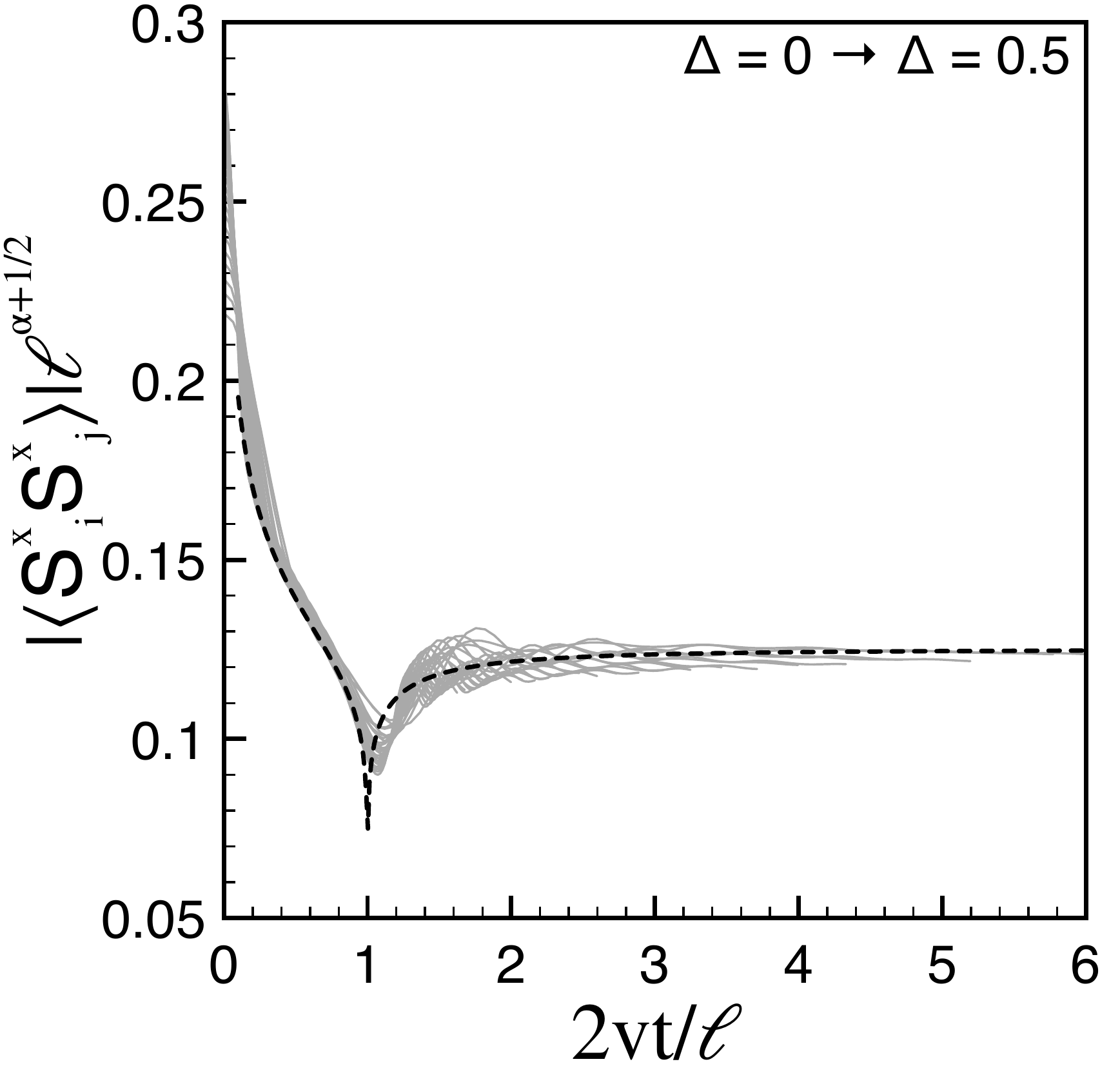}
\includegraphics[width=0.45\textwidth]{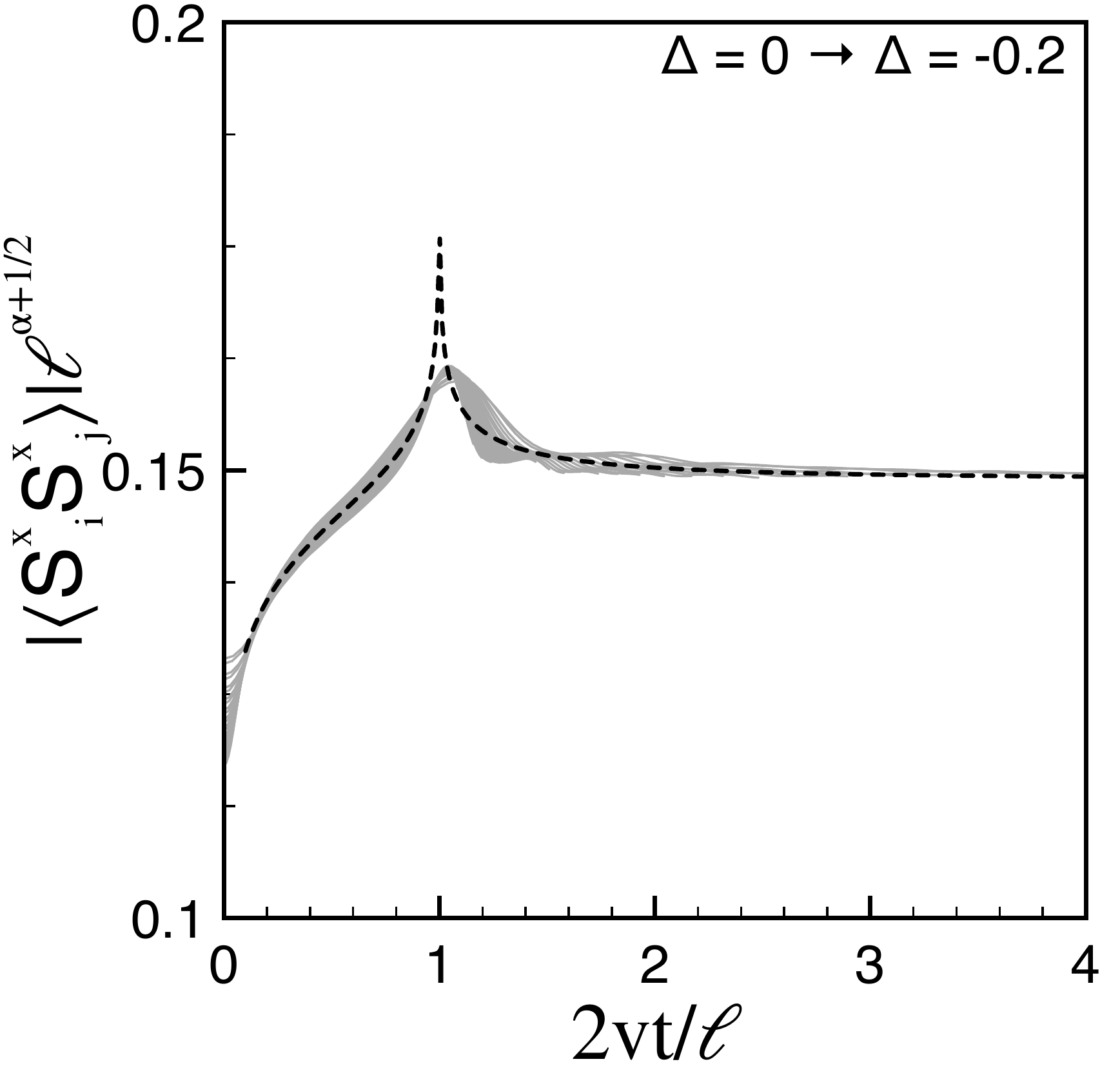}\includegraphics[width=0.45\textwidth]{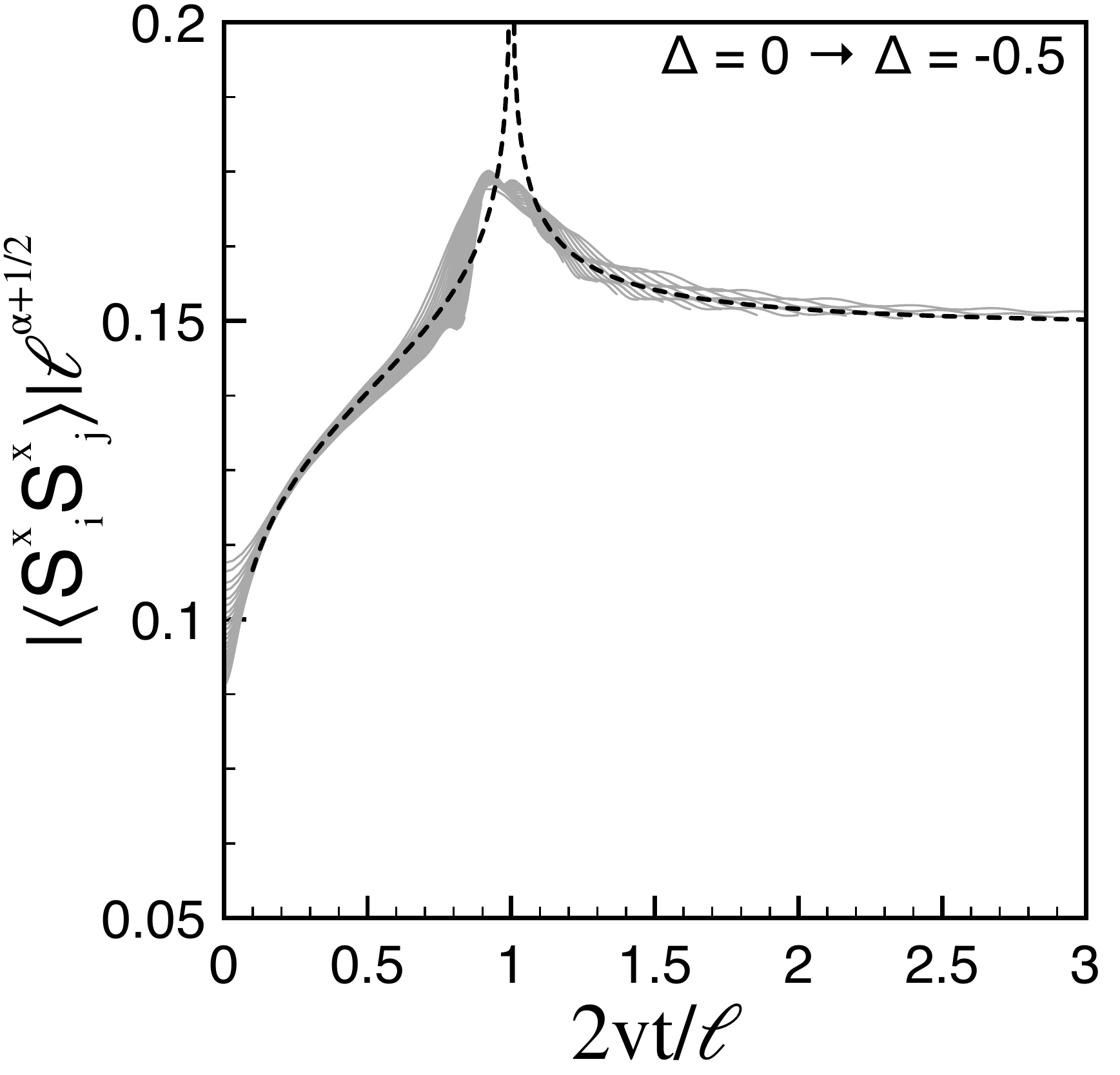}
\caption{\label{figSxSx_vs_t/x} Space-time scaling of 
$|\langle S^{x}_{i}S^{x}_{j}\rangle| \ell^{\alpha+1/2}$ with $\ell = j-i$ for
different values of the post-quench interaction strengths
$\Delta$. After rescaling the correlators, the numerical results (grey
lines) for different distances $\ell$ collapse almost one on top of
one another. The dashed black lines is the universal Luttinger liquid
prediction.}
\end{figure}
%%%%%%%%%%%%%%%%%%%%%%%%%%%%%%%%%%%%%%%%%%

A different way of exhibiting the level of agreement between our numerical
results and the LL predictions is presented in
Fig.~\ref{figSxSx_vs_t/x}, which displays the space-time scaling 
behaviour of the transverse correlation function in terms of the
rescaled variable $2vt/\ell$. The Luttinger liquid result
(\ref{Eq_LM_SxSx}) predicts that correlator for fixed $\Delta$ but
different distances $\ell$ should collapse on top of the same scaling
function, once it has been rescaled by $\ell^{1/2+\alpha}$. Of course,
the usual restrictions $\ell,\, vt, |\ell-2vt| \gg a$, cf. section
~\ref{ssec:Iquench}, continue to apply. Fig. \ref{figSxSx_vs_t/x} 
demonstrates that our numerical results indeed show a very nice
data collapse for all considered quenches. This is consistent with the
good stability of the fitting parameter $A^{x}$ shown in
Figs~\ref{figSxSx_vs_t_1} and  \ref{figSxSx_vs_t_2}. 
We note that the scaling plots in Fig. \ref{figSxSx_vs_t/x} are in
good agreement with the LL prediction even rather close to the light
cone: the smoothed peaks get sharper with increasing $\ell$ and they
get closer and closer to the asymptotic LL results.
The good agreement of our numerical results with the LL prediction is rather
surprising, given that our fits have been performed with a single free
parameter, $A^x$, which moreover turns out to be very close to its
equilibrium value. 

%%%%%%%%%%%%%%%%%%%%%%%%%%%%%%%%%%%%%%%%%%%%
\subsection{Cutoff effects}  
%%%%%%%%%%%%%%%%%%%%%%%%%%%%%%%%%%%%%%%%%%%%
\begin{figure}[t]
\center
\includegraphics[width=0.42\textwidth]{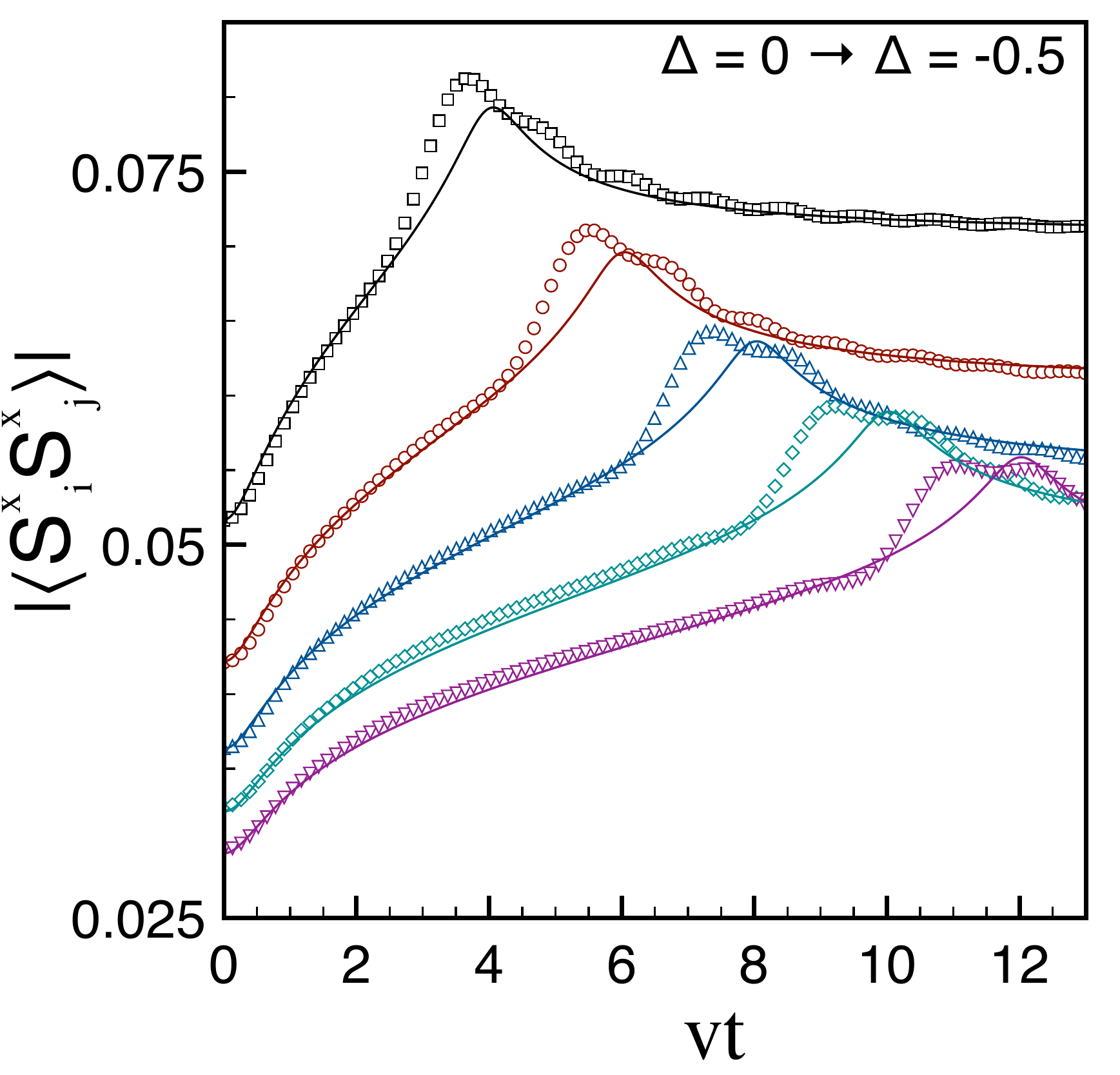} 
\includegraphics[width=0.42\textwidth]{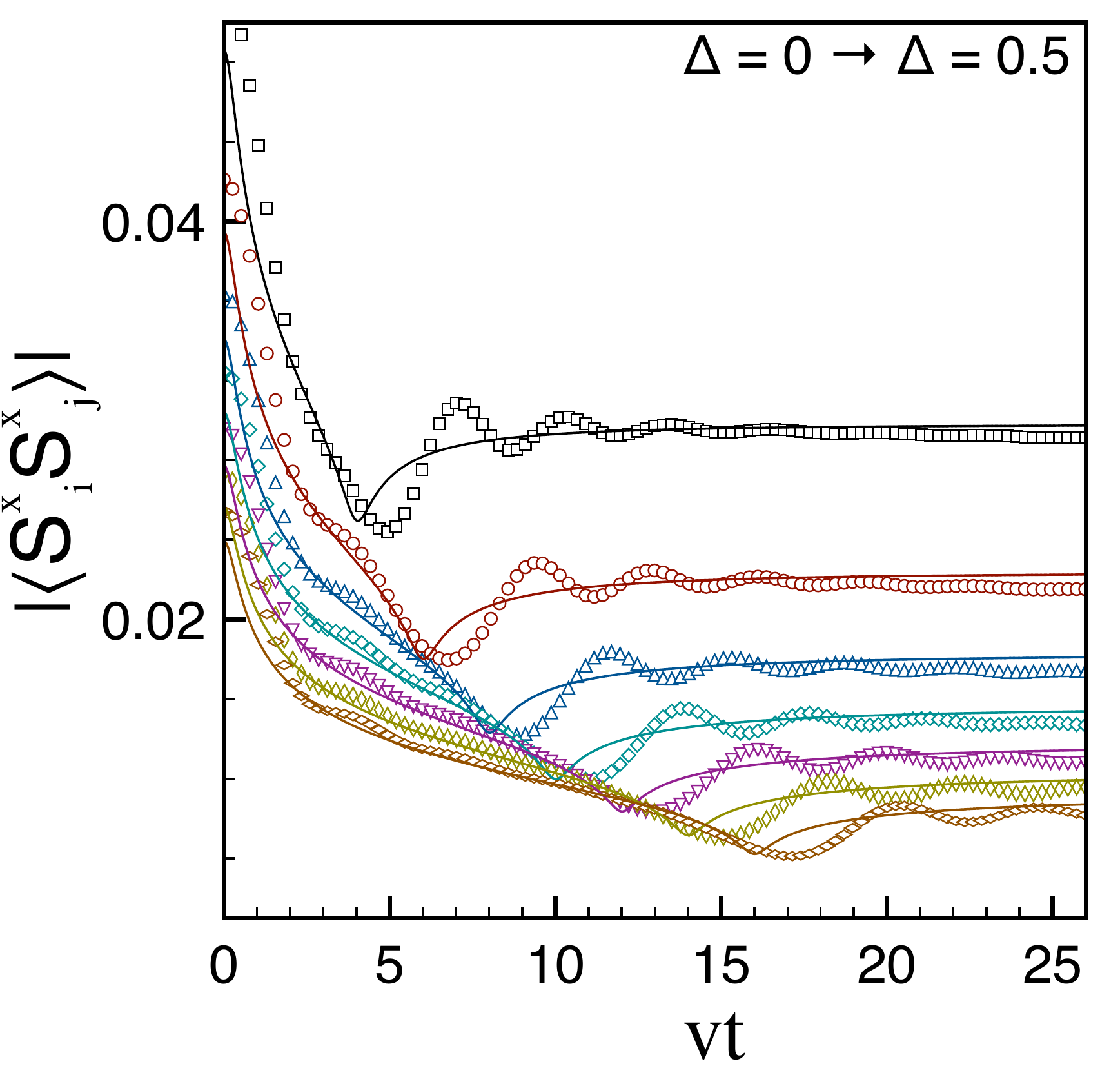}
\caption{\label{Fig:cutoff} Fit of $|\langle
S^{x}_{i}S^{x}_{j}\rangle|$ with $\ell = j-i$ to the LL prediction taking into
account cutoff effects through Eq. (\ref{cutoffF}). The cutoff
improves the agreement with the numerical data for short times, but
not in the vicinity of the light cone. Left panel:
$\ell=8,12,16,20,24$. Right panel: $\ell=8,12,16,20,24,28, 32$.
}
\end{figure}
%%%%%%%%%%%%%%%%%%%%%%%%%%%%%%%%%%%%%%%%%%%%

In the previous section we concluded that Luttinger liquid theory provides
a very good description of the quench dynamics of the XXZ Hamiltonian
as long as $\ell,\, vt, |\ell-2vt| \gg a_0=1$. A natural question
is whether retaining an adjustable cutoff parameter in the LL approach
could lead to an improved description of the short-time dynamics and
the behaviour close to the light cone.
The effect of the cutoff in on the transverse spin correlator is given
by Eq. (\ref{cutoffF}), where $a$ is proportional but not necessarily
equal to $a_0$. Using $a$ as a fit parameter we obtain the results
shown in Fig.~\ref{Fig:cutoff}. We find that, reassuringly, the
amplitude $A^x$ is left unchanged by the introduction of the cutoff.
Inspection of Fig.~\ref{Fig:cutoff} reveals that fitting the value of
the cutoff significantly improves the agreement with the data for
short times, but not close to the light cone. The main remaining difference
between our numerical results and the LL prediction takes the form of a
small oscillatory contribution inside the light cone. This effect goes
beyond the LL approximation, but is small in the transverse
correlation function. This will no longer be the case for the
longitudinal correlation function, to which we turn next.
 
%%%%%%%%%%%%%%%%%%%%%%%%%%%%%%%%%%%%%%%%%%%%%%%%%%%%%%% 
\section{Longitudinal spin correlation functions}
%%%%%%%%%%%%%%%%%%%%%%%%%%%%%%%%%%%%%%%%%%%%%%%%%%%%%%% 
The analysis of the longitudinal correlation function $\langle
S^{z}_{j}S^{z}_{j+\ell}\rangle$ is significantly more involved than that of
$\langle S^{x}_{j}S^{x}_{j+\ell}\rangle$. This is because the LL prediction 
(\ref{Eq_LM_SzSz}) now involves several contributions of similar size,
and because there are substantial effects not captured by simple
Luttinger liquid theory. The latter are strongest close to the light cone
at $t^*=\ell/2v$ and we now turn to their description.
%%%%%%%%%%%%%%%%%%%%%%%%%%%%%%%%%%%%%%%%
\subsection{Vicinity of the light cone}
\begin{figure}[t]
\center
\includegraphics[width=0.45\textwidth]{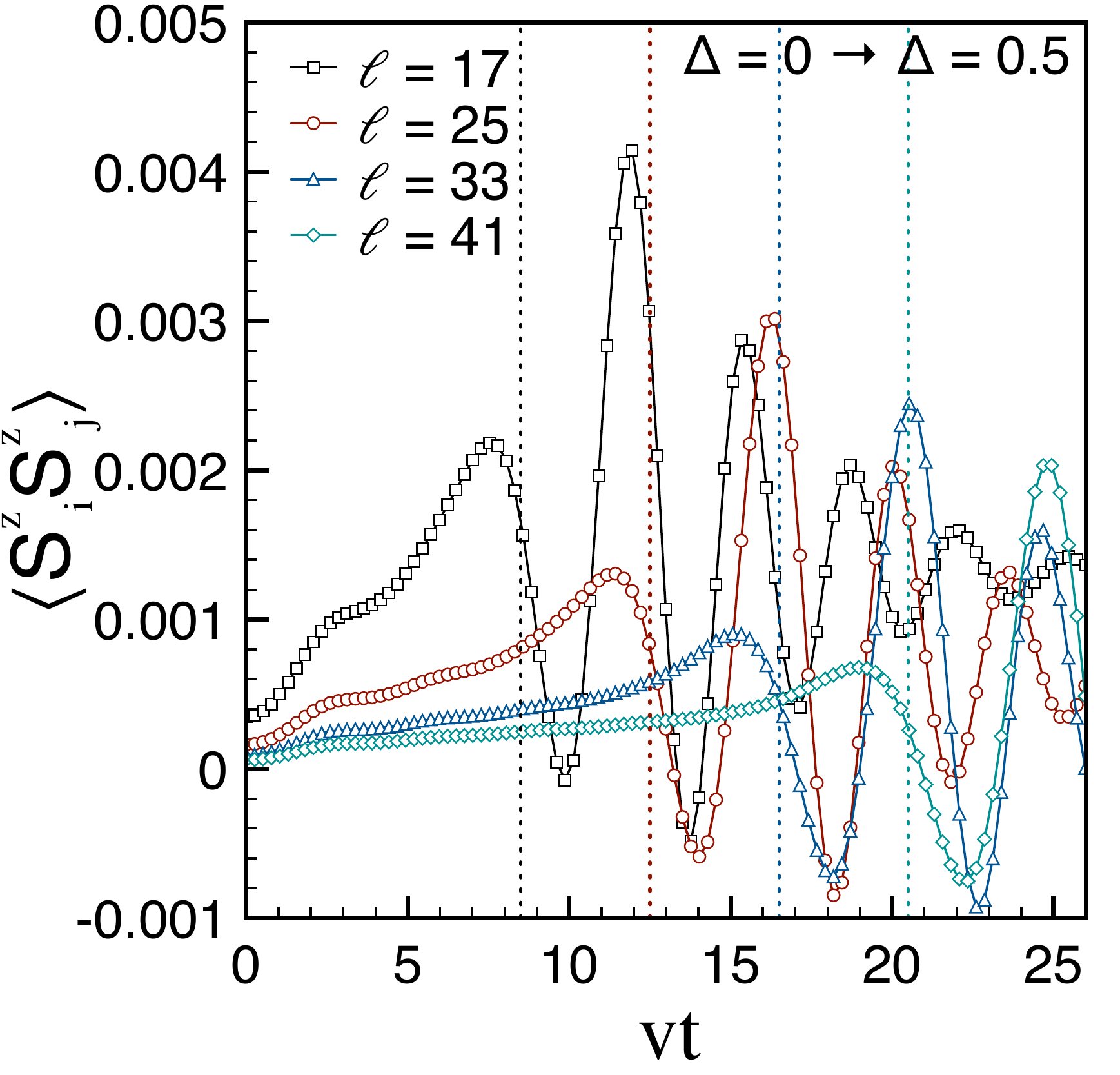} 
\includegraphics[width=0.45\textwidth]{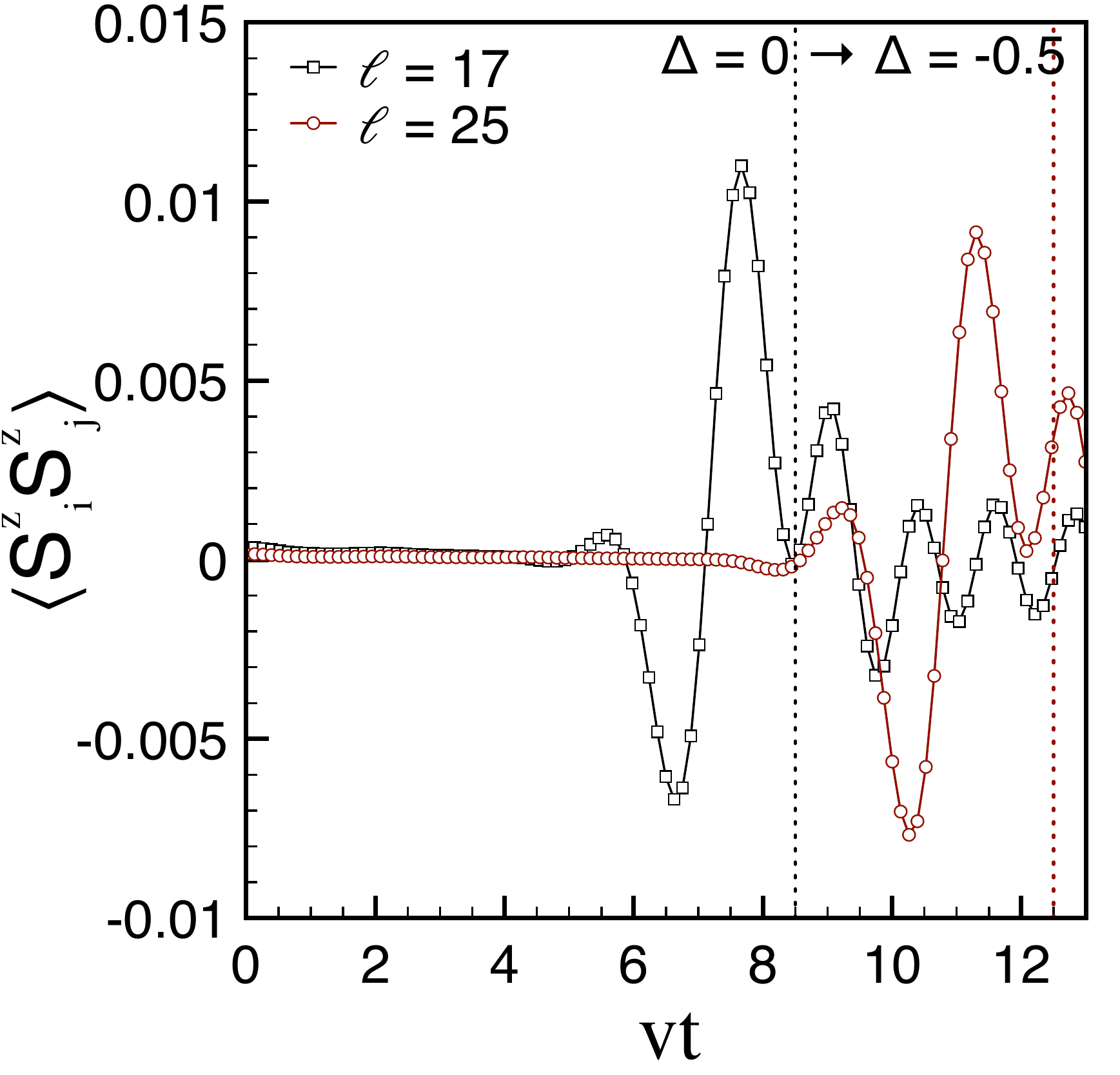} 
\caption{\label{FigSz}
$\langle S^{z}_{i}S^{z}_{j}\rangle$ as function of time for
several separations $\ell=j-i$. Large oscillations appear when the light
cone (represented by a vertical line of the same colour as the data)
is approached.
}
\end{figure}

%%%%%%%%%%%%%%%%%%%%%%%%%%%%%%%%%%%%%%%%
In Fig. \ref{FigSz} we show some typical results for quenches to
positive (left panel) and negative (right panel) values of $\Delta$. 
 %%%%%%%%%%%%%%%%%%% FIGURE CONE %%%%%%%%%%%%%%%%%%%%%%%
 %%%%%%%%%%%%%%%%%%%%%%%%%%%%%%%%%%%%%%%%%%%%%%%%%%

We fix the distance $\ell$ to be an odd
integer in order to account for a strong even/odd effect present in
the initial correlations
\begin{equation}
\langle S^{z}_{j}S^{z}_{j+\ell}\rangle\Big|_{t=0} = \frac{(-1)^{\ell}-1}{2\pi^2\ell^2}.
\end{equation}
We note that this effect is captured by the LL approximation
(\ref{Eq_LM_SzSz}), since the smooth and the staggered terms
(proportional respectively to $B^{z}$ and $A^{z}$) are very close in
magnitude. Inspection of Fig. \ref{FigSz} shows that as the light cone
(represented by dashed vertical lines in the figure) is approached for
a given $\ell$, large oscillations in time ensue. This behaviour is
plainly not encoded in the Luttinger liquid approximation and can be
understood qualitatively as follows. In the Luttinger liquid
approximation the longitudinal spin correlation function exhibits
a strong (quadratic) singularity at the light cone. The behaviour in
this regime is determined by the high-energy modes, i.e. the
ultraviolet part of the spectrum. However, at high energies the
Luttinger liquid approximation is no longer expected to provide a good
description of the Heisenberg model, as lattice effects (such as saddle
points of the spinon dispersion) are important. On the other hand, one
may hope that Luttinger liquid approximation is applicable
sufficiently far away from the light cone. In order to investigate
this possibility it is useful to consider the short-time ($2vt\leq
\ell$) and the long-time ($2vt\geq \ell$) behaviour separately.

%%%%%%%%%%%%%%%%%%%%%%%%%%%%%%%%%%%%%%%%
\subsection{Space-time scaling behaviour}
%%%%%%%%%%%%%%%%%%%%%%%%%%%%%%%%%%%%%%%%
Before starting the quantitative analysis, we discuss the space-time
scaling of this correlator. In the space-time scaling regime with
$\ell,t\to\infty$ with their ratio fixed, the staggered and the smooth
terms scale differently with the latter going like $\ell^{-2}$ while
the former like $\ell^{-2-\beta}$ with $\beta=K^2-1$. Given the dependence
of $K$ on $\Delta$, cf. Eq. (\ref{KvBA}), the staggered term is
leading for $\Delta>0$ while the smooth contribution dominates for
$\Delta<0$. 
 %%%%%%%%%%%% FIGURE SzSz   vs 2vt/x  %%%%%%%%%%%%%
\begin{figure}[t]
\center
\includegraphics[width=0.45\textwidth]{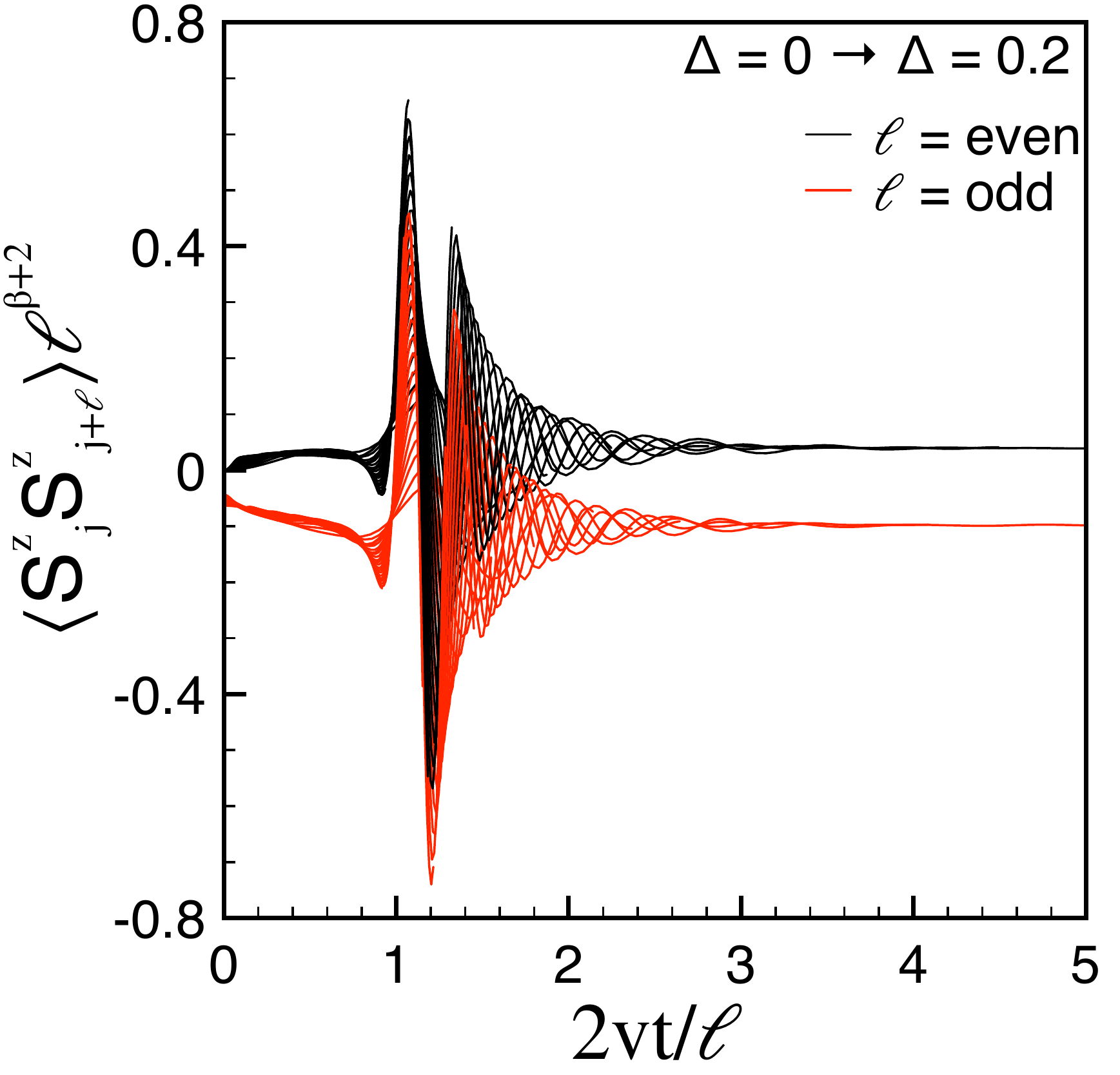}
\includegraphics[width=0.45\textwidth]{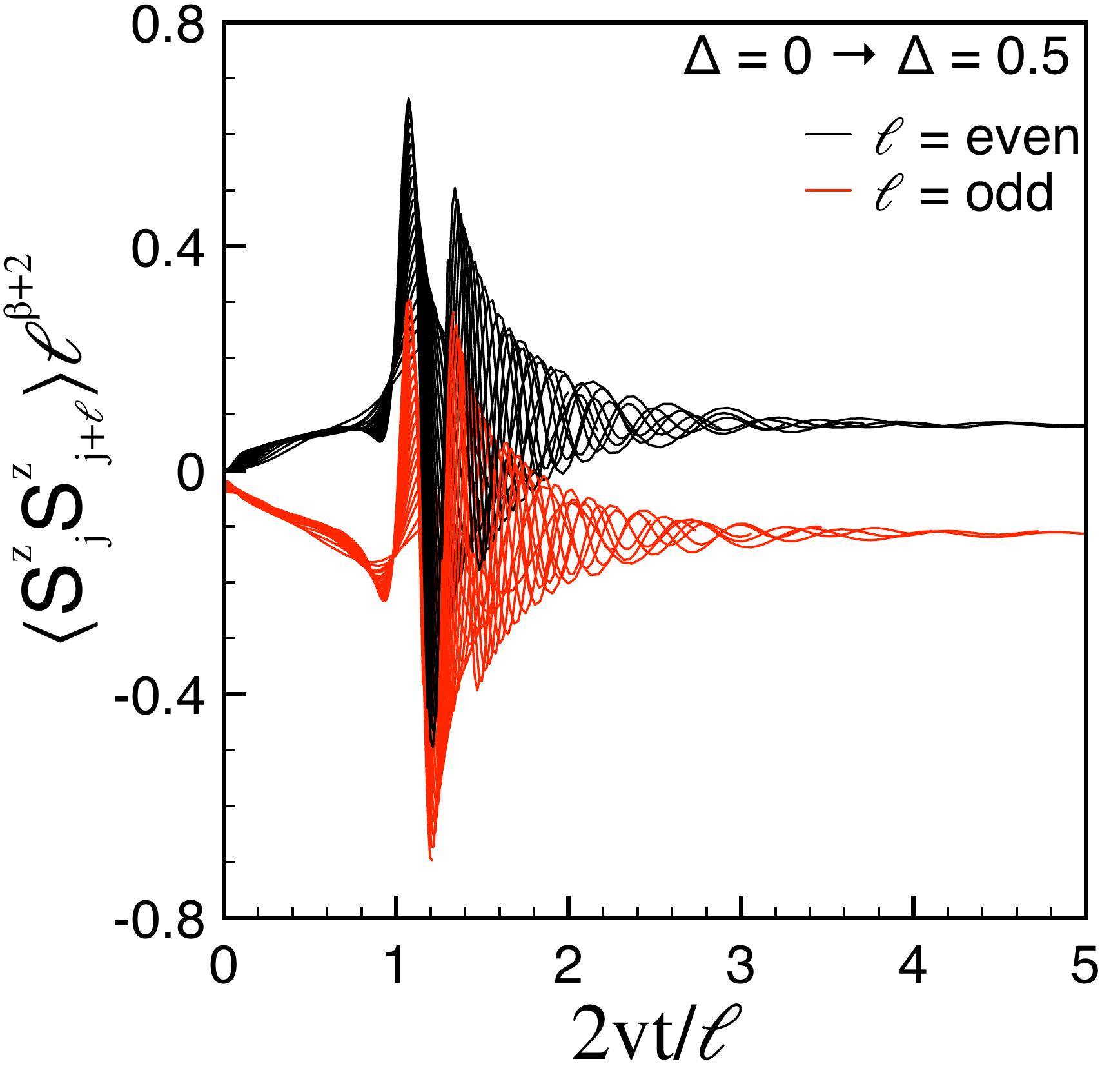}
\caption{\label{figSzSz_vs_t/x_1_rep} 
Space-time scaling of  $\langle S^{z}_{j}S^{z}_{j+\ell}\rangle$ for
$\Delta>0$. The numerical results (black and red lines) for different
distances $\ell$ generate two regular envelopes associated with even
and odd $\ell$ respectively.
}
\end{figure}
We first consider the repulsive regime $\Delta>0$. In
Fig.~\ref{figSzSz_vs_t/x_1_rep} we show a space-time scaling plot of
$\langle S^{z}_{j}S^{z}_{j+\ell}\rangle \ell^{\beta+2}$ as a function
of the scaled variable $2vt/\ell$ for several values of $\ell$.
The most striking feature of the longitudinal correlations are their
pronounced enhancement in the vicinity of the light cone, as compared
to their initial values (in absolute terms the correlations are still
small due to the factor of $\ell^{\beta+2}$). This is quite different
from behaviour seen in the transverse spin-spin correlation function.

For a given parity of $\ell$ the numerical results are seen to exhibit
a fair data collapse outside the light cone. The absence of symmetry
around zero has its origin in the presence of both staggered and
smooth contributions to the correlation function. In the vicinity of
the light cone as well as in its interior the oscillations described
above spoil any data collapse, but the different curves form nice,
regular envelopes. 

\begin{figure}[t]
\center
\includegraphics[width=0.45\textwidth]{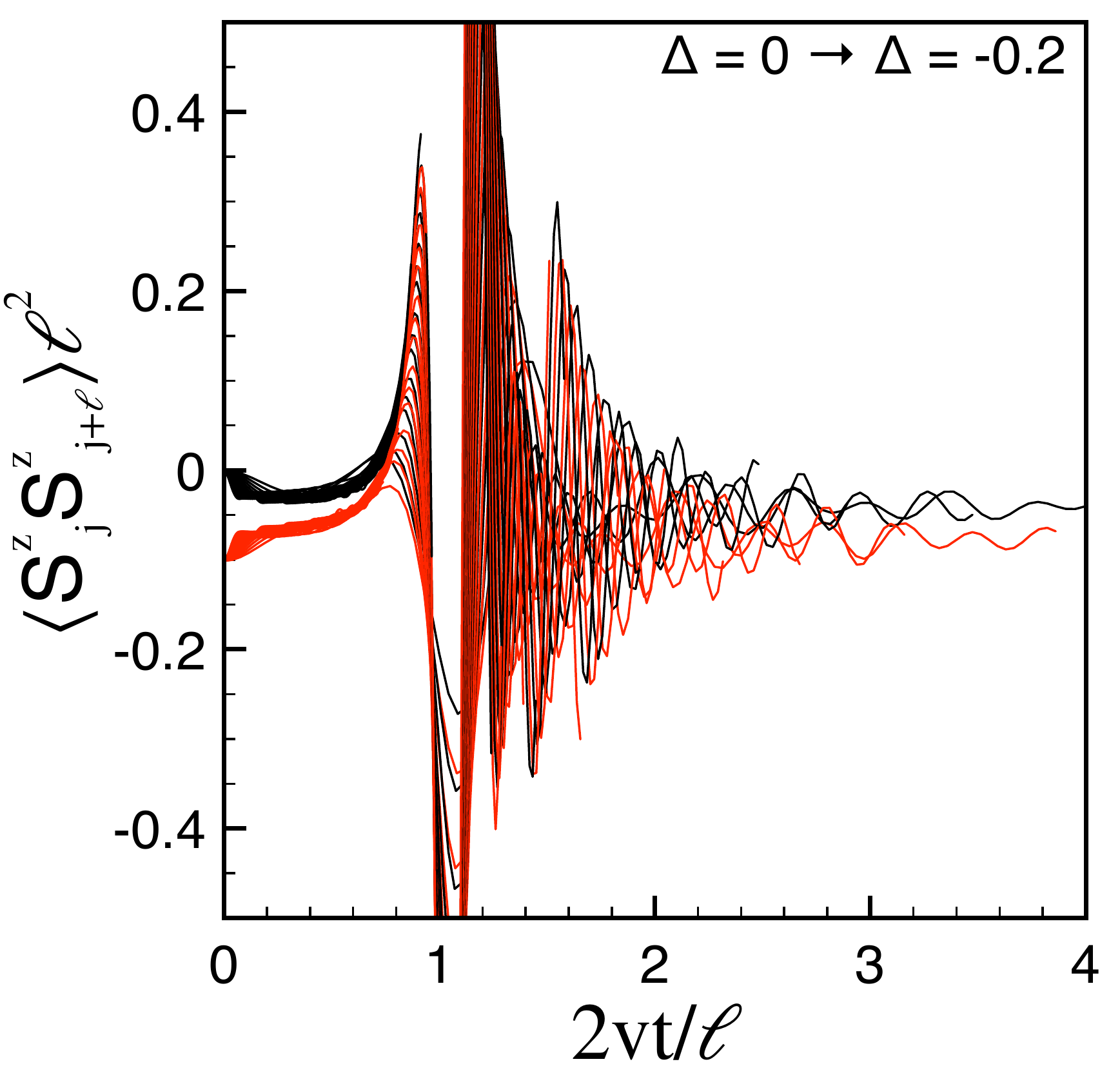}
\includegraphics[width=0.45\textwidth]{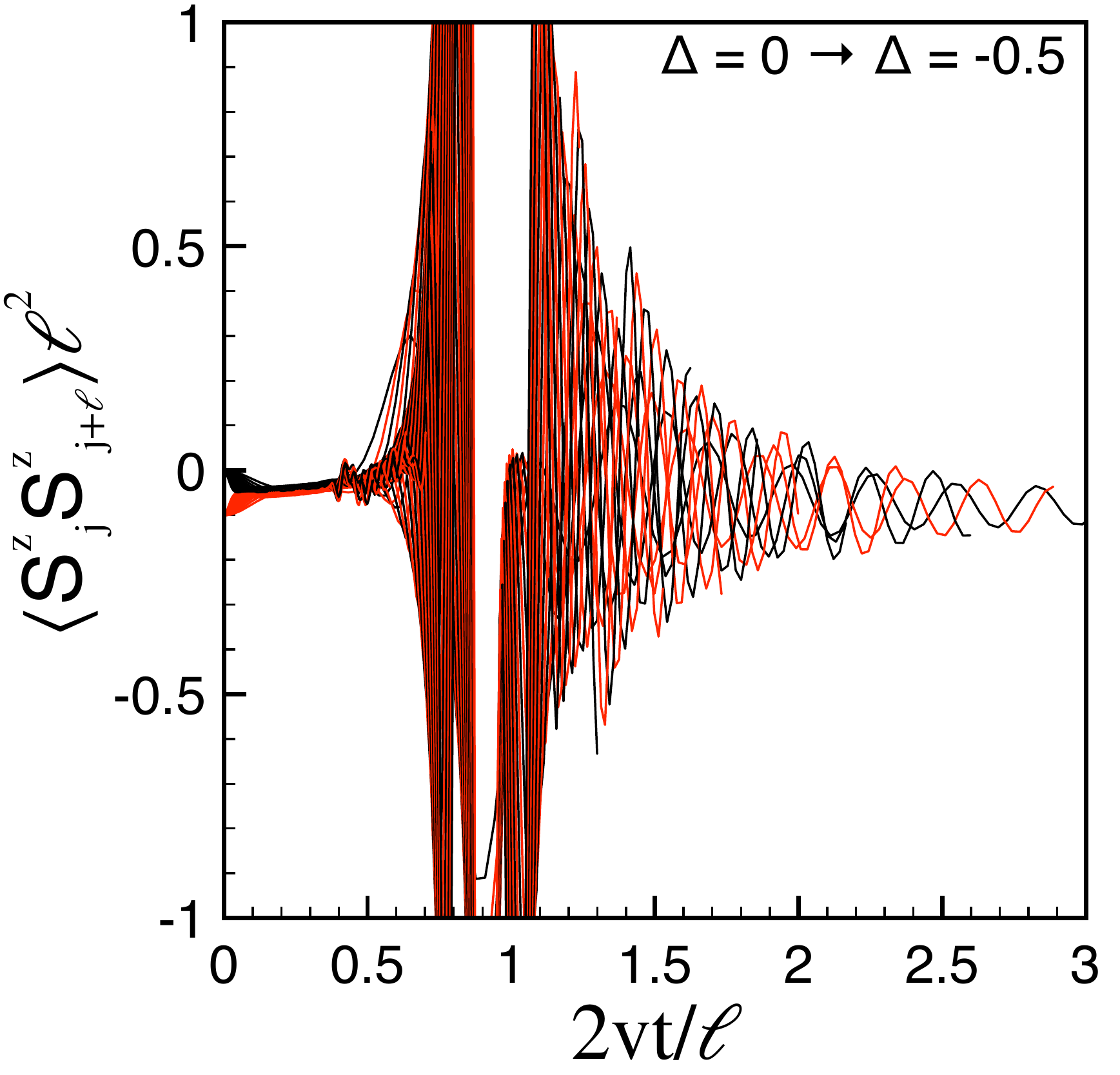}
\caption{\label{figSzSz_vs_t/x_1} 
Space-time scaling of  $\langle S^{z}_{j}S^{z}_{j+\ell}\rangle$ for
$\Delta<0$. We observe fair data collapse at short times, and
irregular envelopes at late times.
}
\end{figure}
In the attractive regime we consider the scaling behaviour of $\langle
S^{z}_{j}S^{z}_{j+\ell}\rangle \ell^{2}$. As is shown in
Fig. \ref{figSzSz_vs_t/x_1}, there again is a very strong enhancement
of longitudinal correlations in the vicinity of the light cone. 
There is a good data collapse only for short times and large
oscillations start playing an important role at an early stage of the
time evolution (for $\Delta =-0.5$ the oscillations start at $2vt/\ell
\simeq 0.4$). Moreover, we observe that the oscillations are much more
irregular than for $\Delta>0$.

%%%%%%%%%%%%%%%%%%%%%%%%%%%%%%%%%%%%%%%%
\subsection{Behaviour outside the light cone $2vt\leq \ell$}
\begin{figure}[ht]
\center
\includegraphics[width=0.4\textwidth]{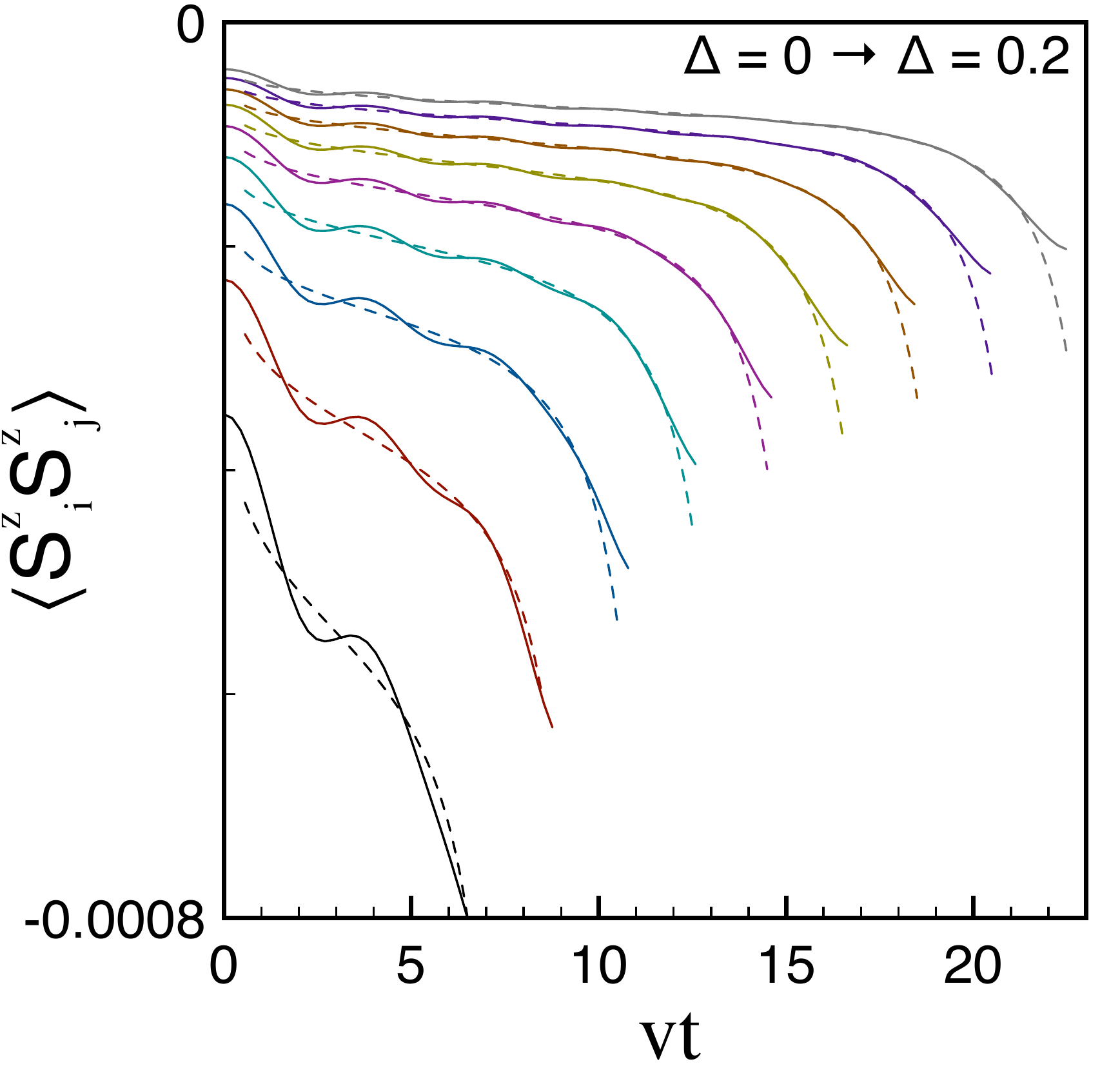}
\includegraphics[width=0.4\textwidth]{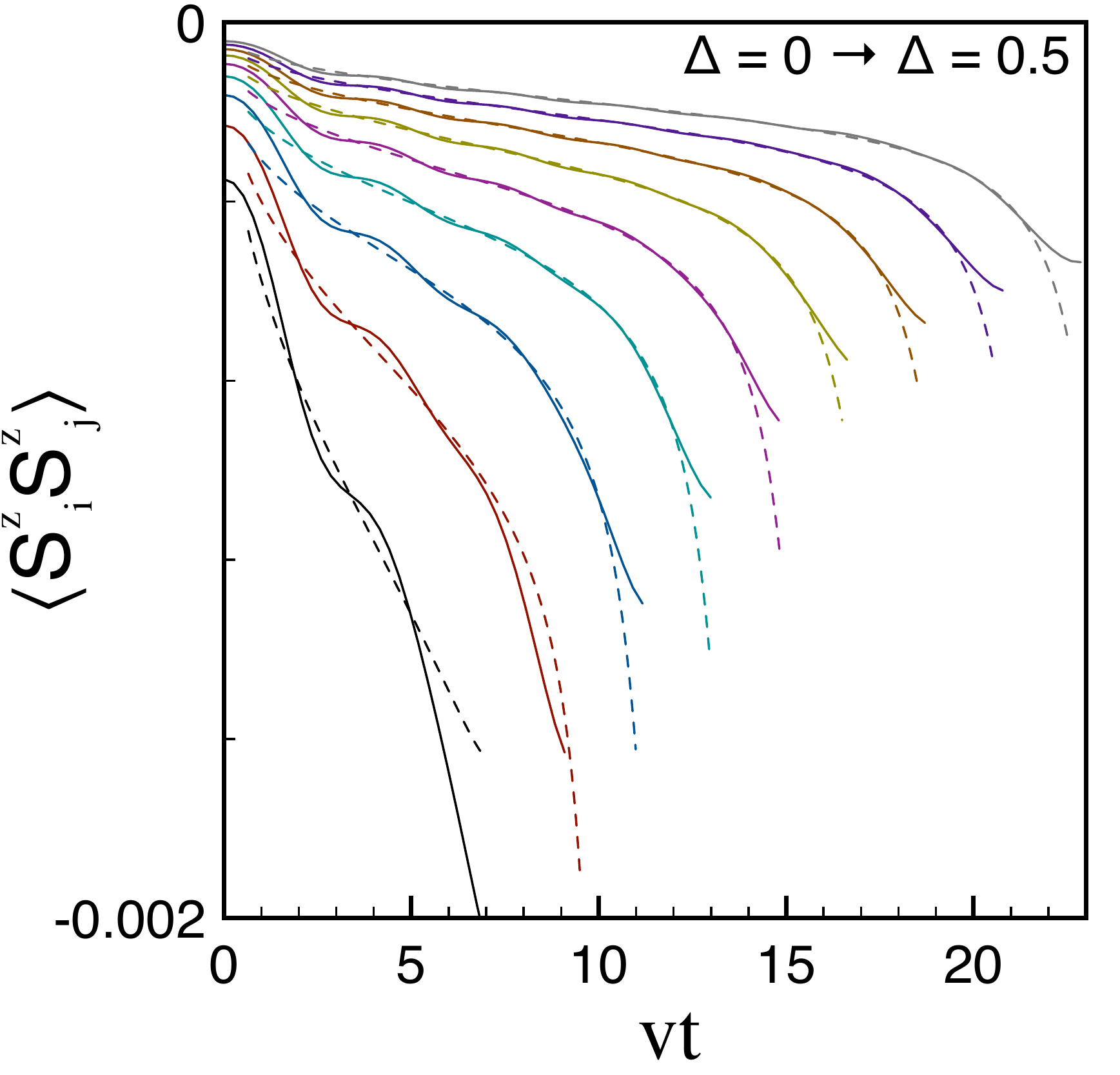}\\
\includegraphics[width=0.4\textwidth]{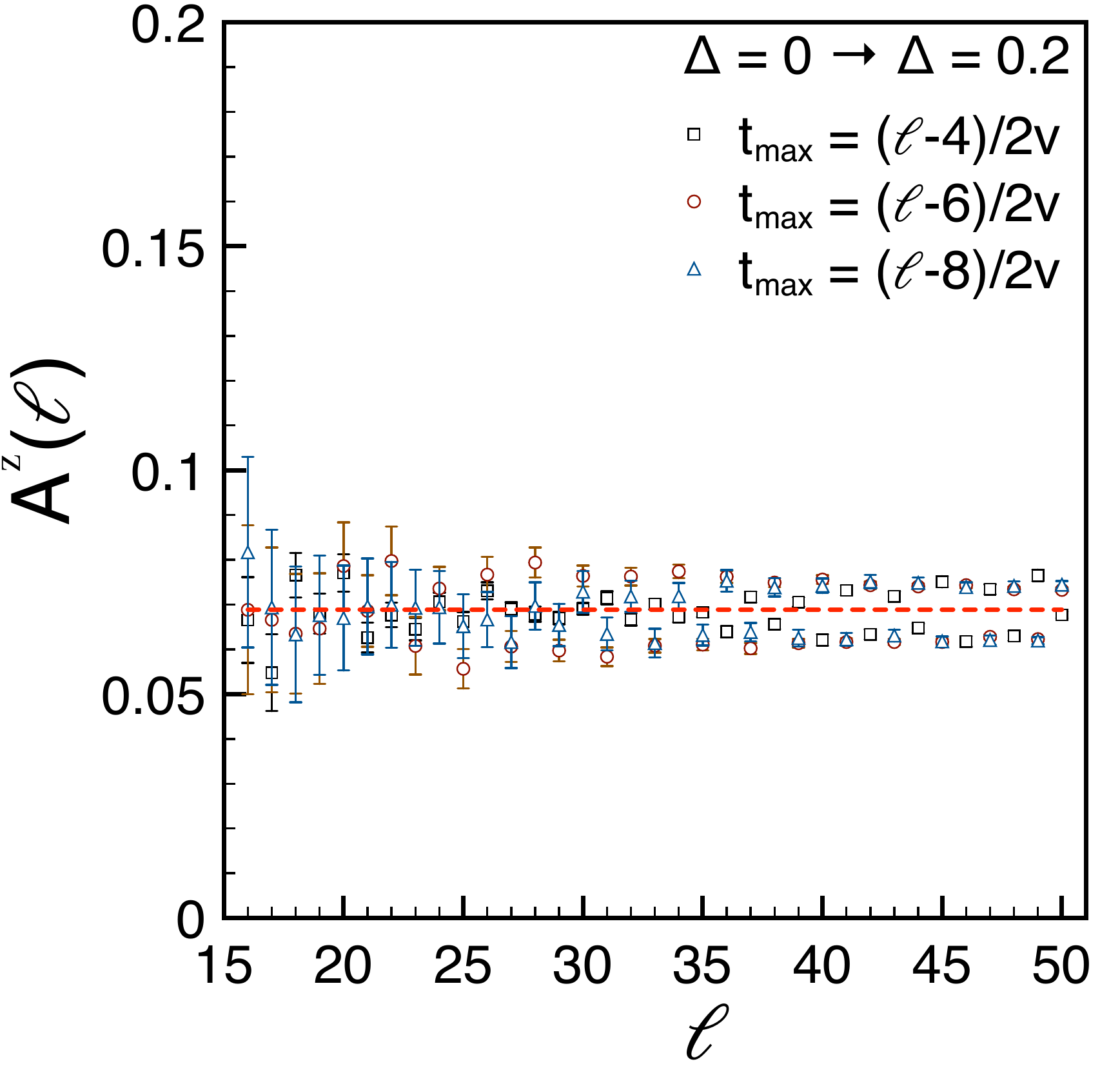}
\includegraphics[width=0.4\textwidth]{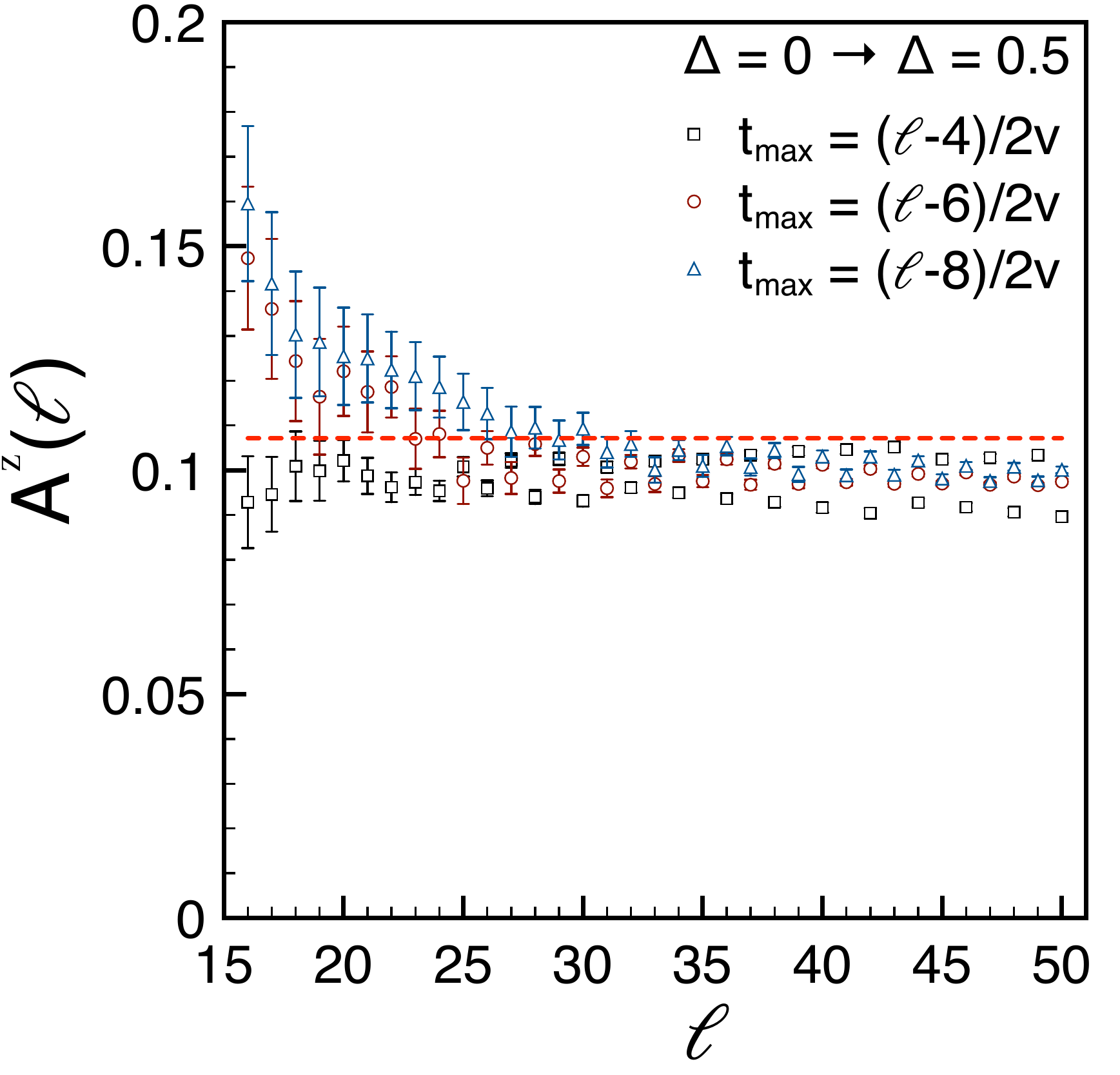}\\
\includegraphics[width=0.4\textwidth]{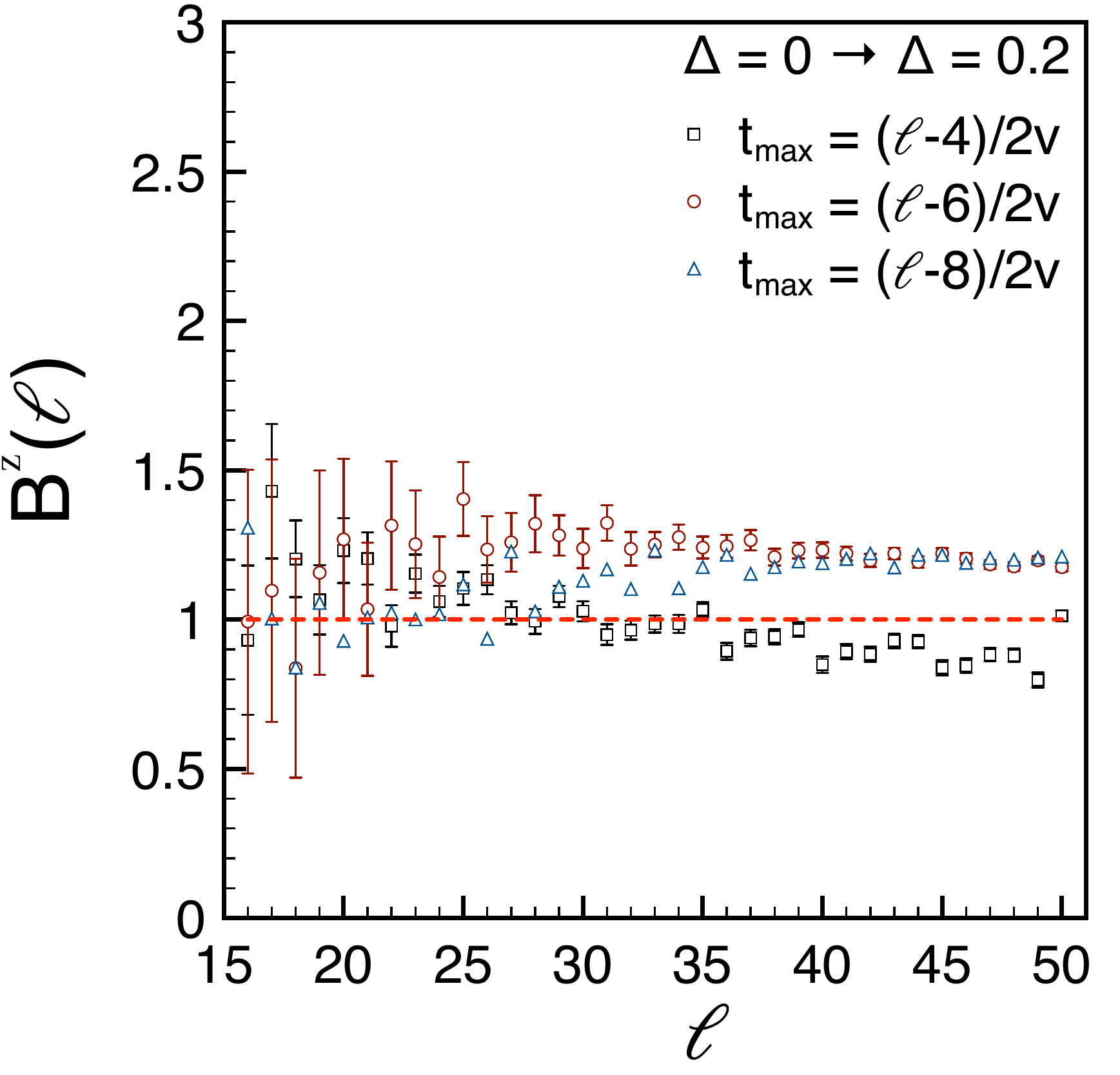}
\includegraphics[width=0.4\textwidth]{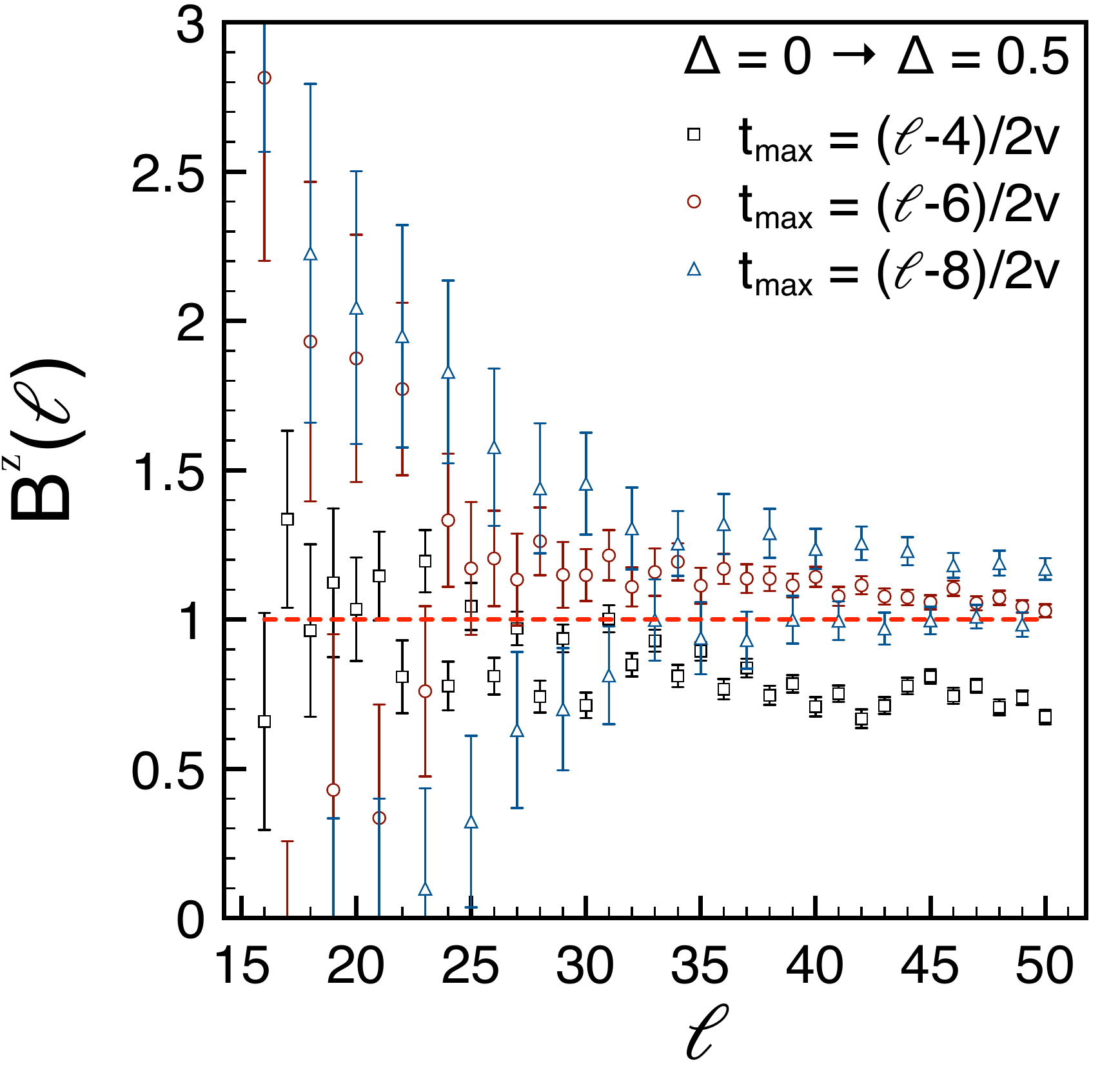}
\caption{\label{figSzSz_vs_t_short}
(Top) Short time regime of $\langle S^{z}_{i}S^{z}_{j}\rangle$, for
two different interaction strengths in the repulsive regime $\Delta>0$
and distances $\ell =j-i = 17,\, 21,\, 25,\, \dots ,\, 49$. The
numerical data are shown as full lines, while dashed lines represent
best fits with the LL approximation (fitting with $t_{max} = (\ell-6)/2 v$).  
(Centre/Bottom) Amplitudes $A^{z}$ and $B^{z}$ obtained from best
fits with different values of $t_{max}$. The dashed red lines are the
equilibrium values of the amplitude, i.e. $A^{z}_{0} = 1$, $A^{z}_{1}
= 0.0689$ (for $\Delta = 0.2$) and  $A^{z}_{1} = 0.1071$ (for $\Delta
= 0.5$).}  
\end{figure}

\begin{figure}[ht]
\center
\includegraphics[width=0.4\textwidth]{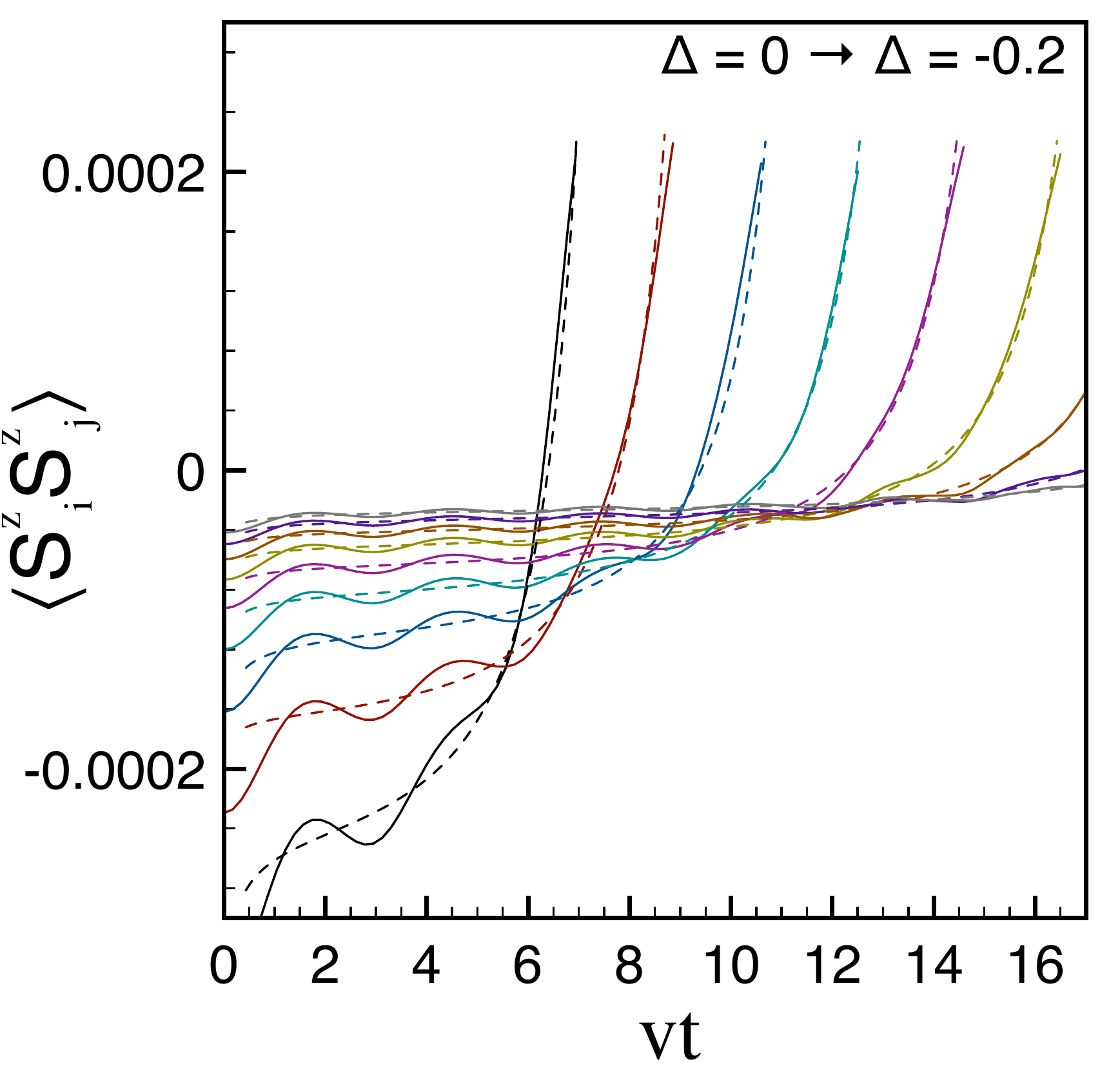}
\includegraphics[width=0.4\textwidth]{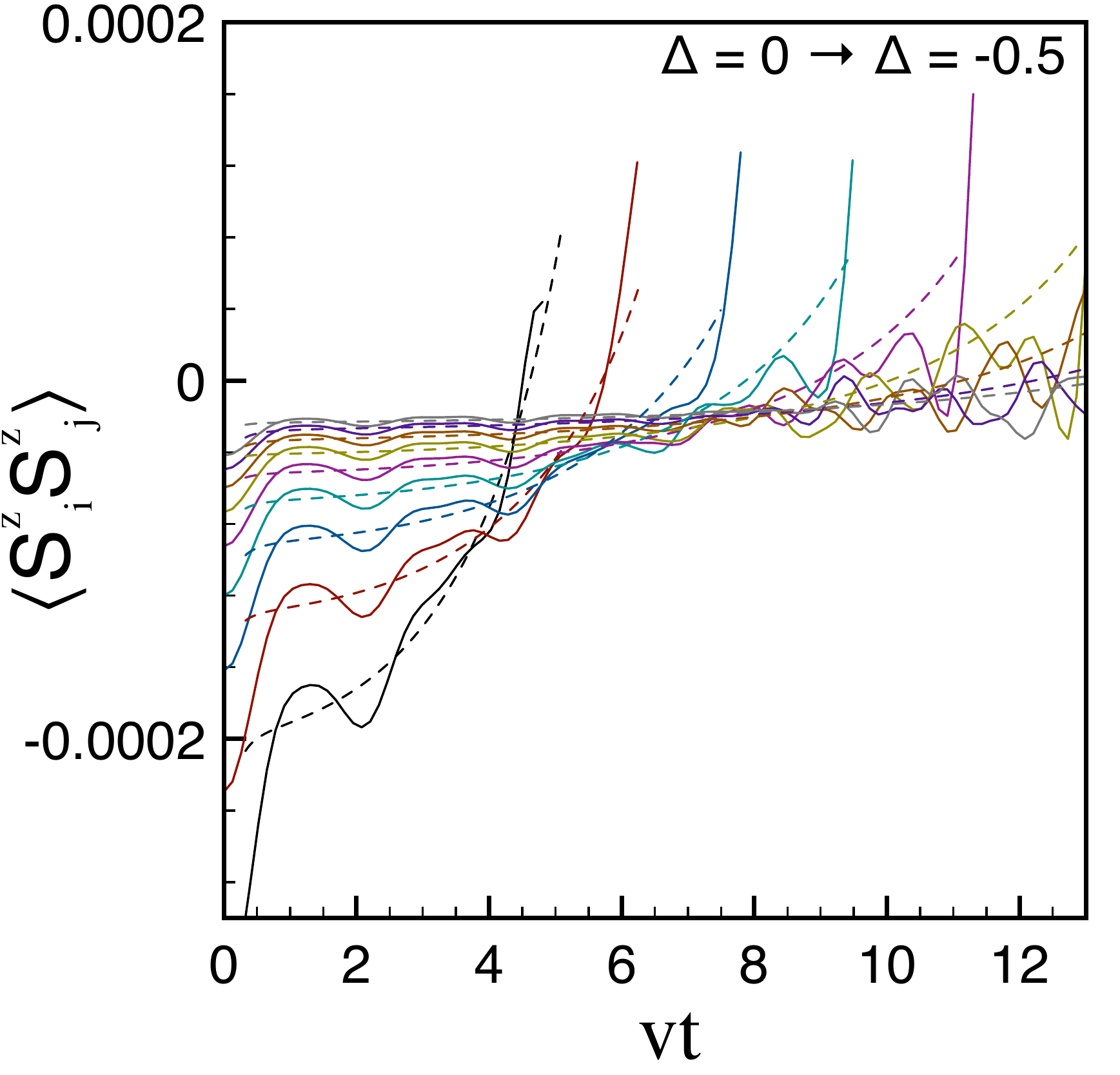}\\
\includegraphics[width=0.4\textwidth]{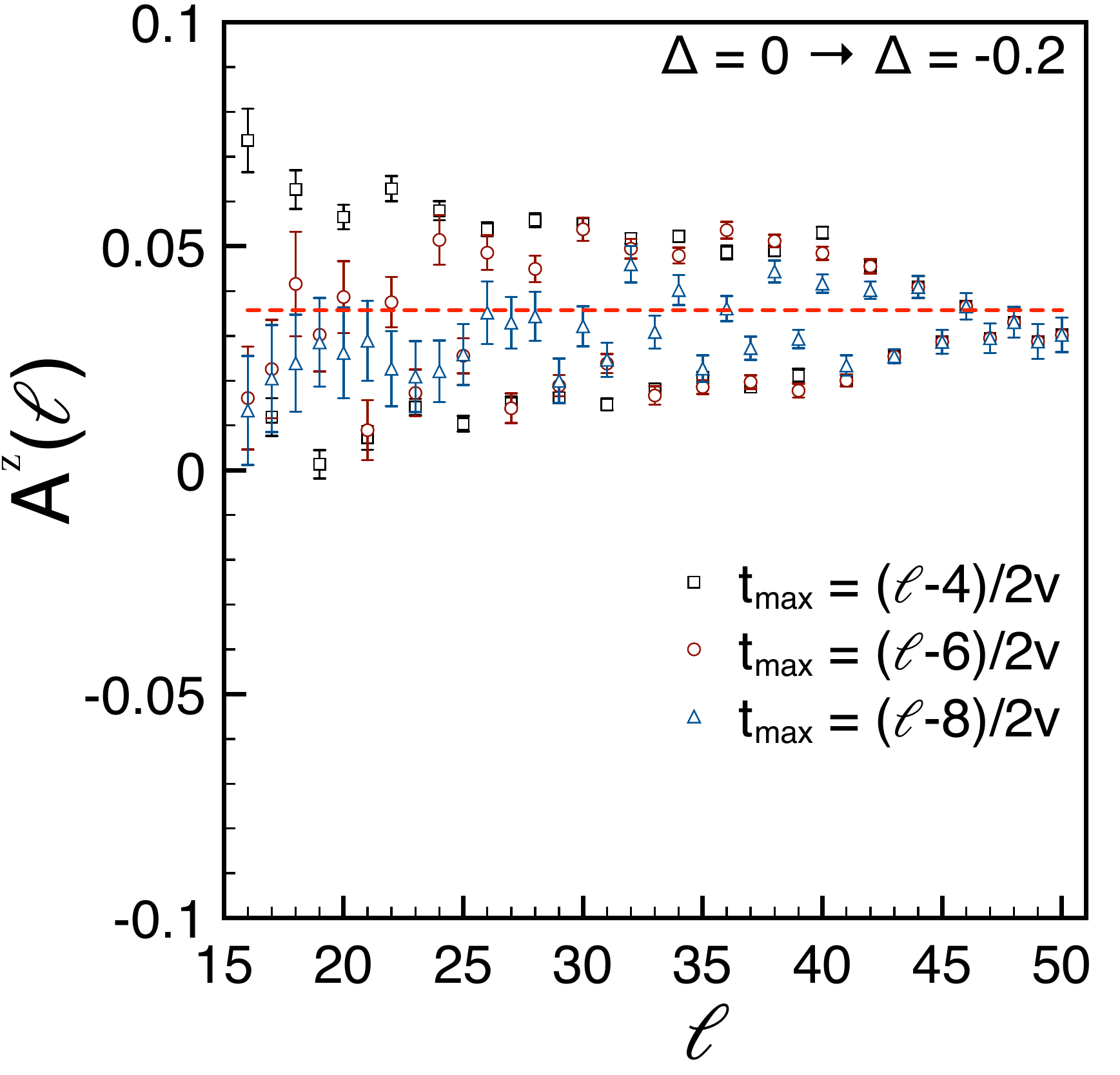}
\includegraphics[width=0.4\textwidth]{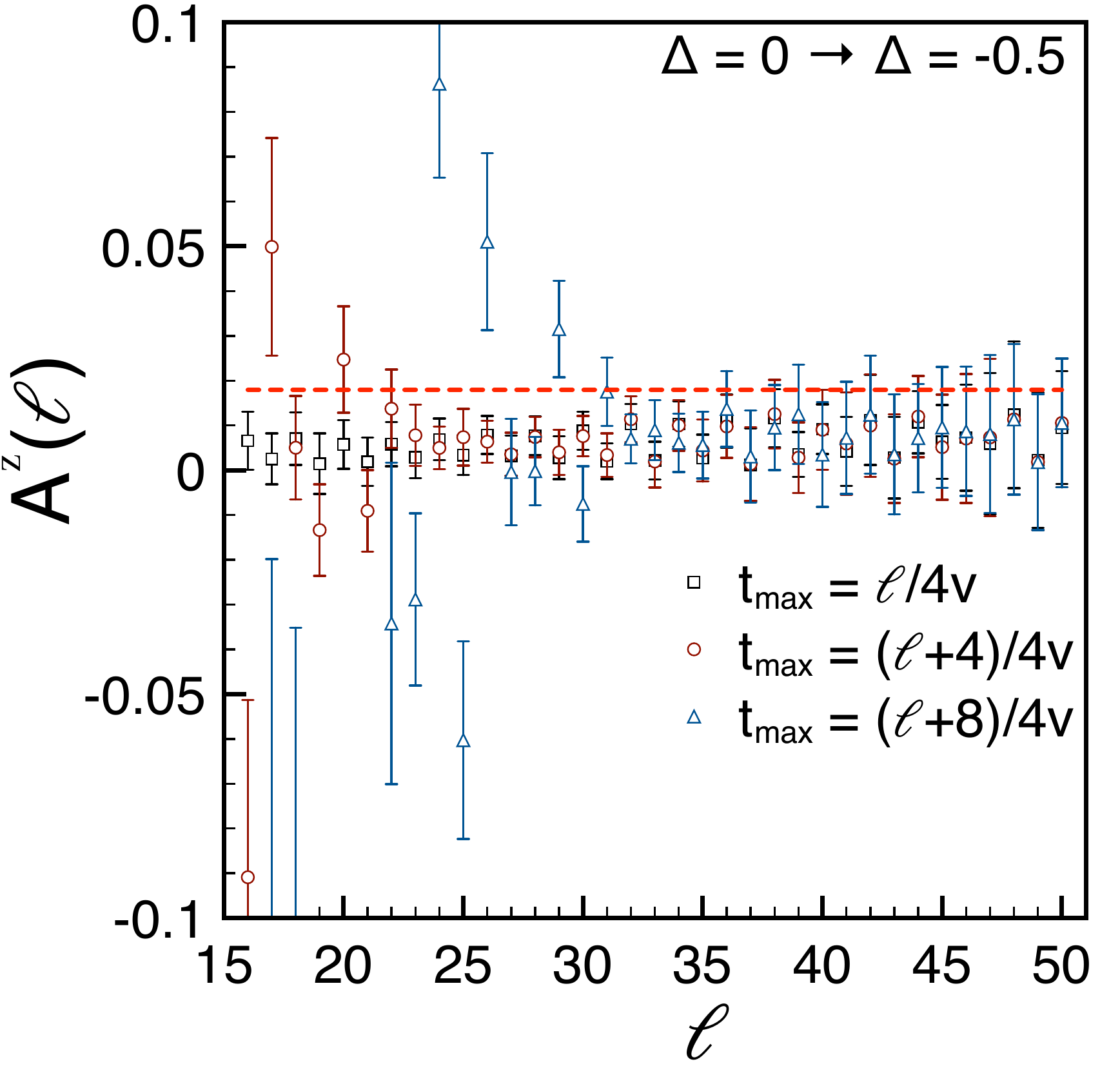}\\
\includegraphics[width=0.4\textwidth]{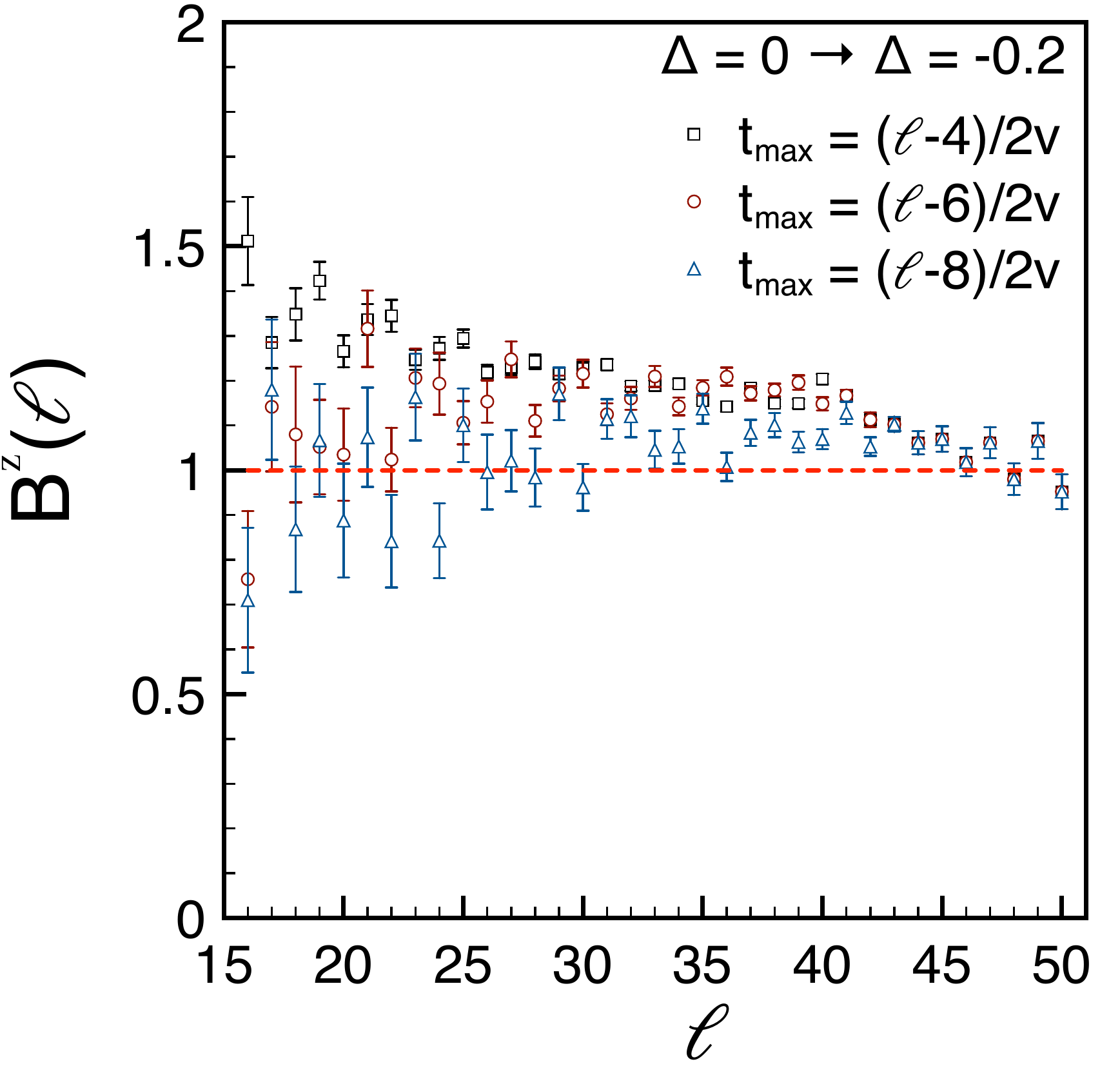}
\includegraphics[width=0.4\textwidth]{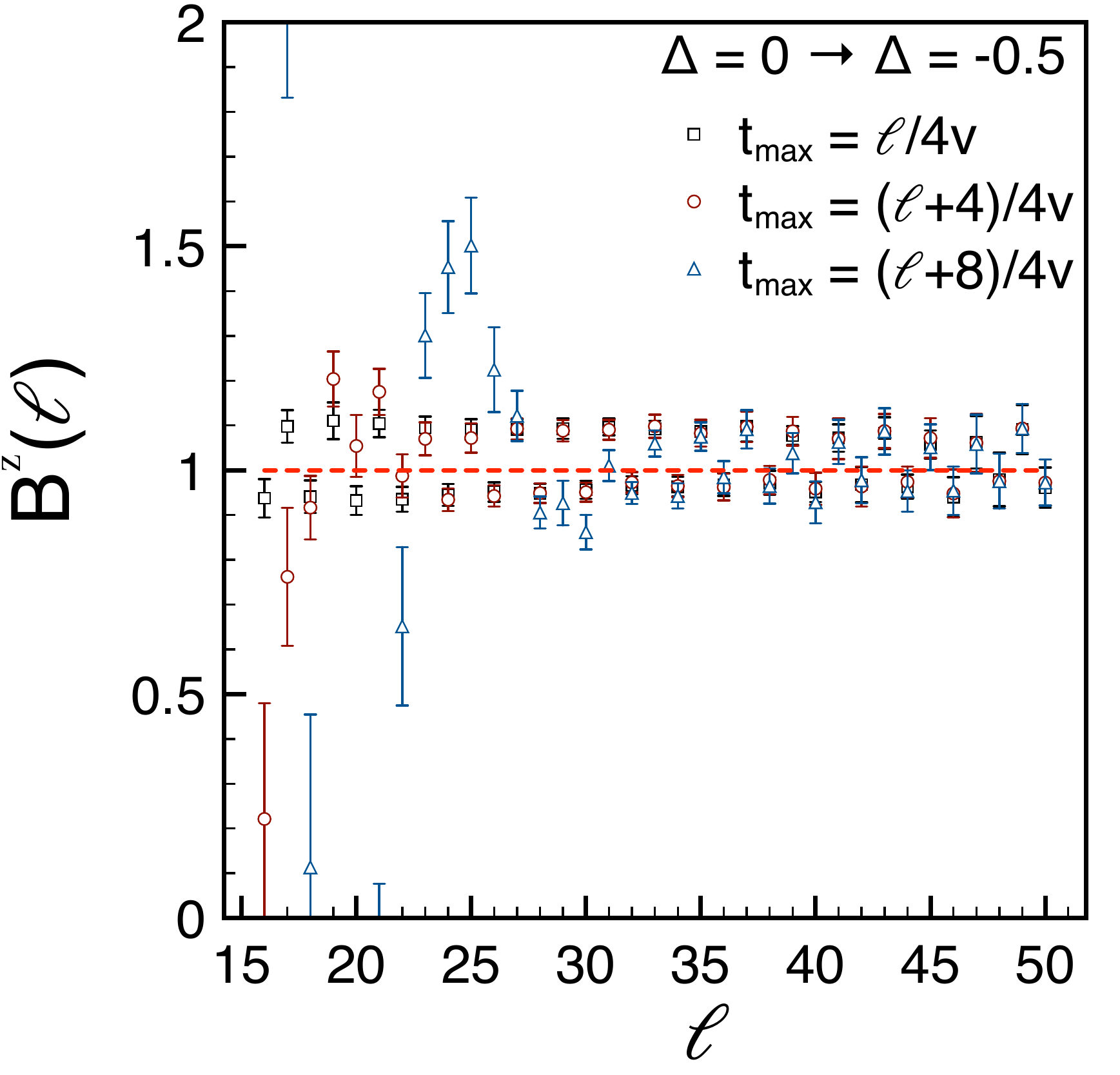}
\caption{\label{figSzSz_vs_t_short2}
(Top) Short time regime of $\langle S^{z}_{i}S^{z}_{j}\rangle$ for
two interaction strengths in the attractive regime $\Delta<0$ and
distances $\ell =j-i = 17,\, 21,\, 25,\, \dots ,\, 49$. 
The numerical data are shown as full lines, while dashed lines represent
best fits with the LL approximation (fitting with $t_{max} =
(\ell-6)/2 v$ for $\Delta = -0.2$ and with $t_{max} = \ell/4 v$ for
$\Delta = -0.5$). 
(Centre/Bottom) Amplitudes $A^{z}$ and $B^{z}$ obtained from fits with
different $t_{max}$. The dashed red lines are the equilibrium values of
the amplitude, i.e. $A^{z}_{0} = 1$, $A^{z}_{1} = 0.0357$ (for $\Delta
= -0.2$)  and  $A^{z}_{1} = 0.0179$ (for $\Delta = -0.5$).} 
\end{figure}

%%%%%%%%%%%%%%%%%%%%%%%%%%%%%%%%%%%%%%%%
We now turn to a quantitative analysis of the longitudinal
correlations outside the light cone, i.e. at short times
$t\leq \ell/2v$. The time dependence of $\langle
S^{z}_{j}S^{z}_{j+\ell}\rangle$ is shown in the top two panels of
Fig.~\ref{figSzSz_vs_t_short} for quenches to $\Delta=0.2$,
$\Delta=0.5$ and in Fig.~ \ref{figSzSz_vs_t_short2} for quenches
to $\Delta=-0.2$, $\Delta=-0.5$. The numerical results are compared to
best fits to the Luttinger liquid prediction (\ref{Eq_LM_SzSz}), which
are shown as dashed lines. The agreement is seen to be satisfactory.
The fitted values for the amplitudes $A^z$ and $B^z$ are shown
in the middle and bottoms panels of Figs~\ref{figSzSz_vs_t_short} 
and ~\ref{figSzSz_vs_t_short2} respectively.

%%%%%%%%%%%% FIGURE SzSz   vs t (short)  Delta > 0 %%%%%%%%%%%%%
%%%%%%%%%%%%%%%%%%%%%%%%%%%%%%%%%%%%%%%%%%%%%%%

%%%%%%%%%%%% FIGURE SzSz   vs t (short)  Delta < 0 %%%%%%%%%%%%%
%%%%%%%%%%%%%%%%%%%%%%%%%%%%%%%%%%%%%%%%%%%%%%%

Some comments on our fit procedure are in order. To avoid the strong
light-cone singularity in the LL approximation at $\ell=2vt$, we
performed fits in the time window $[1,(\ell-\delta\ell)/2v]$ for
several values of $\delta\ell$. These are shown in the bottom four
panels of Figs~\ref{figSzSz_vs_t_short} and  \ref{figSzSz_vs_t_short2}). 
We observe a quite satisfactory stability of the coefficients $A^{z}$
and $B^z$ against changing the time interval over which the fits are
performed. This stability improves with increasing $\ell$, i.e. gets
better in the regime where the LL approximation is expected to work
best. We observe a small even/odd effect in the fitted values of
$A^z$ and $B^z$, which we attribute to subleading terms within
Luttinger liquid theory.
The fitted values of both $A^z$ and $B^z$ are approximately the same as in
equilibrium. In Table \ref{tab2} we report some rough estimates of the
maximal relative differences.

%%%%%%%%%%%%%%%%%%%%%%%%%%%%%%%%%%%%%%%%%%%%%%%%%%%%%
\subsection{LL approximation close to the light cone}
\begin{figure}[t]
\center
\includegraphics[width=0.45\textwidth]{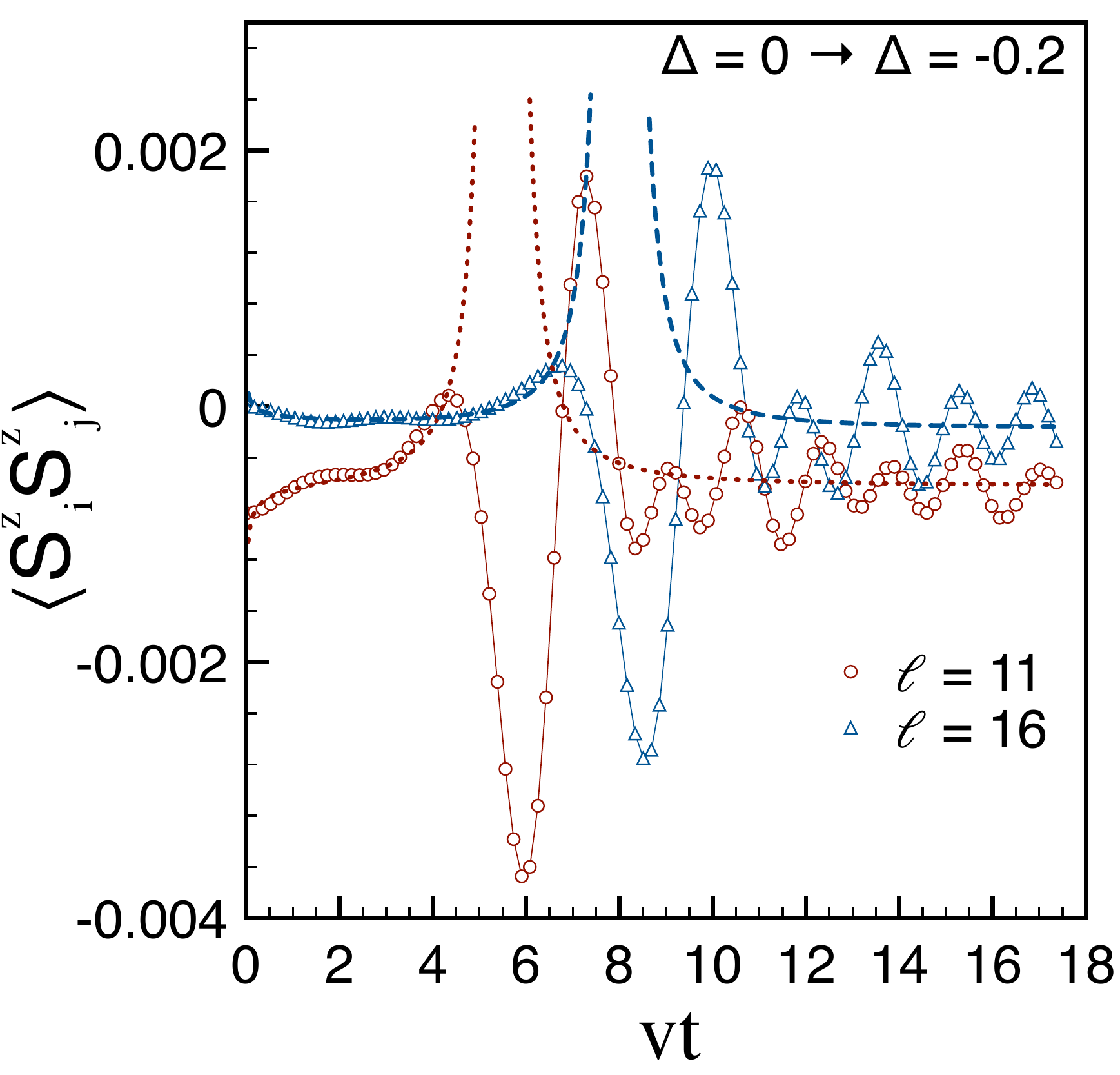} 
\includegraphics[width=0.45\textwidth]{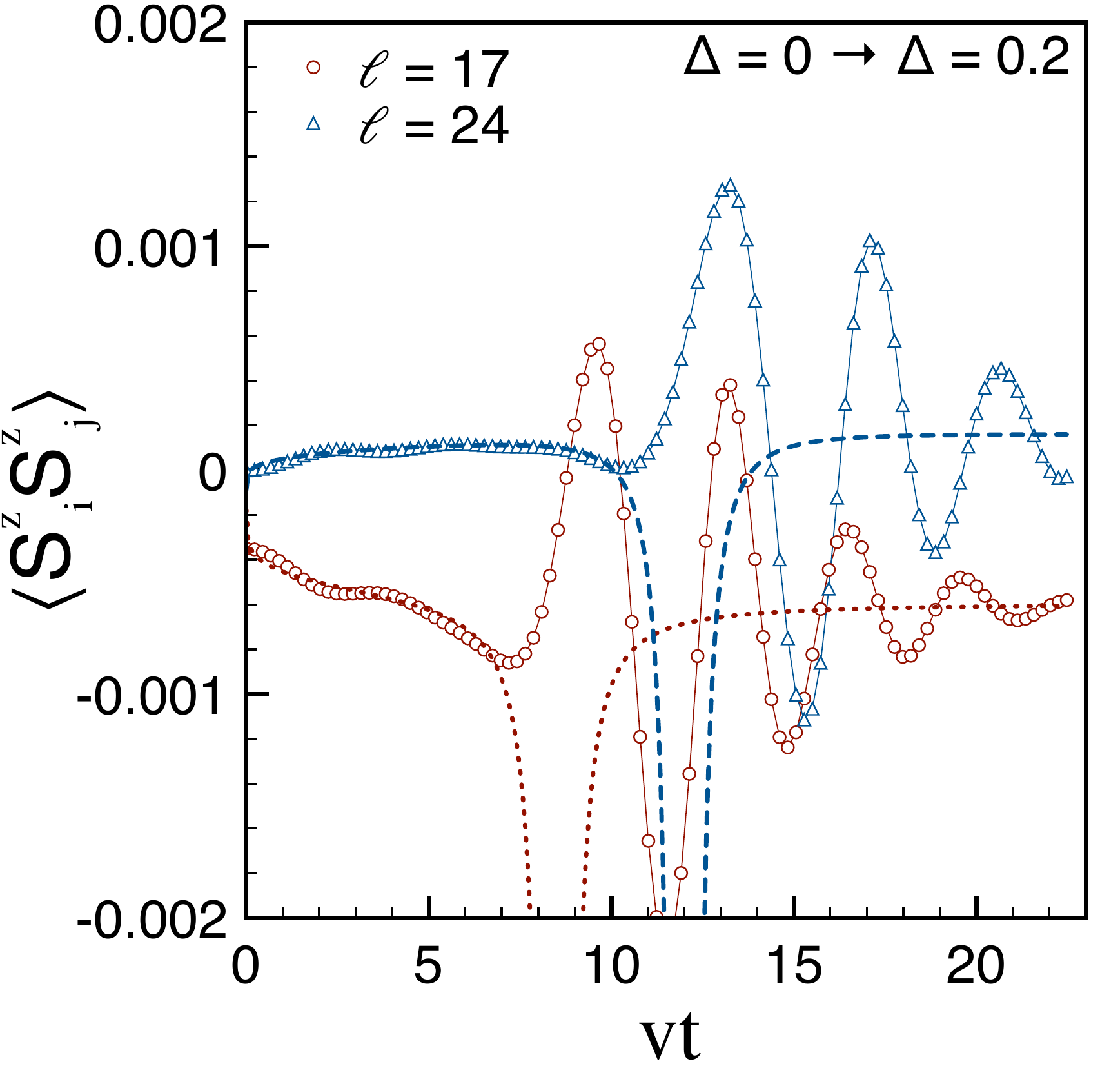} 
\caption{\label{figSzSz_vs_BA}
Numerical results for  $\langle S^{z}_{i}S^{z}_{j}\rangle$ 
compared (for several distances $\ell=j-i$) to the LL approximation with amplitudes fixed by fitting only
short times. While the oscillatory behaviour is not captured by the LL
approximation, it appears to correctly account for the decay of
correlations in the accessible time window.}
\end{figure}

%%%%%%%%%%%%%%%%%%%%%%%%%%%%%%%%%%%%%%%%%%%%%%%%%%%%%
A natural question is to what extent the LL approximation, with
amplitudes $A^z$, $B^z$ determined above, continues to describe the
correlation function close to, and inside, the light cone. In
Fig.~\ref{figSzSz_vs_BA} we present results of such comparisons for
two different quenches. The correlation function is seen to exhibit
decaying oscillatory behaviour inside the light cone. It is evident
that the LL approximation fails to capture the strong
oscillations. However, in all cases the oscillations appear to be
centered around the LL approximation and to decay towards the latter.

%%%%%%%%%%%%%%%%%% FIGURE SzSz vs  LM+BA %%%%%%%%%%%%%%%%
%%%%%%%%%%%%%%%%%%%%%%%%%%%%%%%%%%%%%%%%%%%%%%%

%%%%%%%%%%%%%%%%%%%%%%%%%%%%%%%%%%%%%%%%
\subsection{Physics beyond the LL approximation inside the light cone} 
\begin{figure}[ht]
\center
\includegraphics[width=0.45\textwidth]{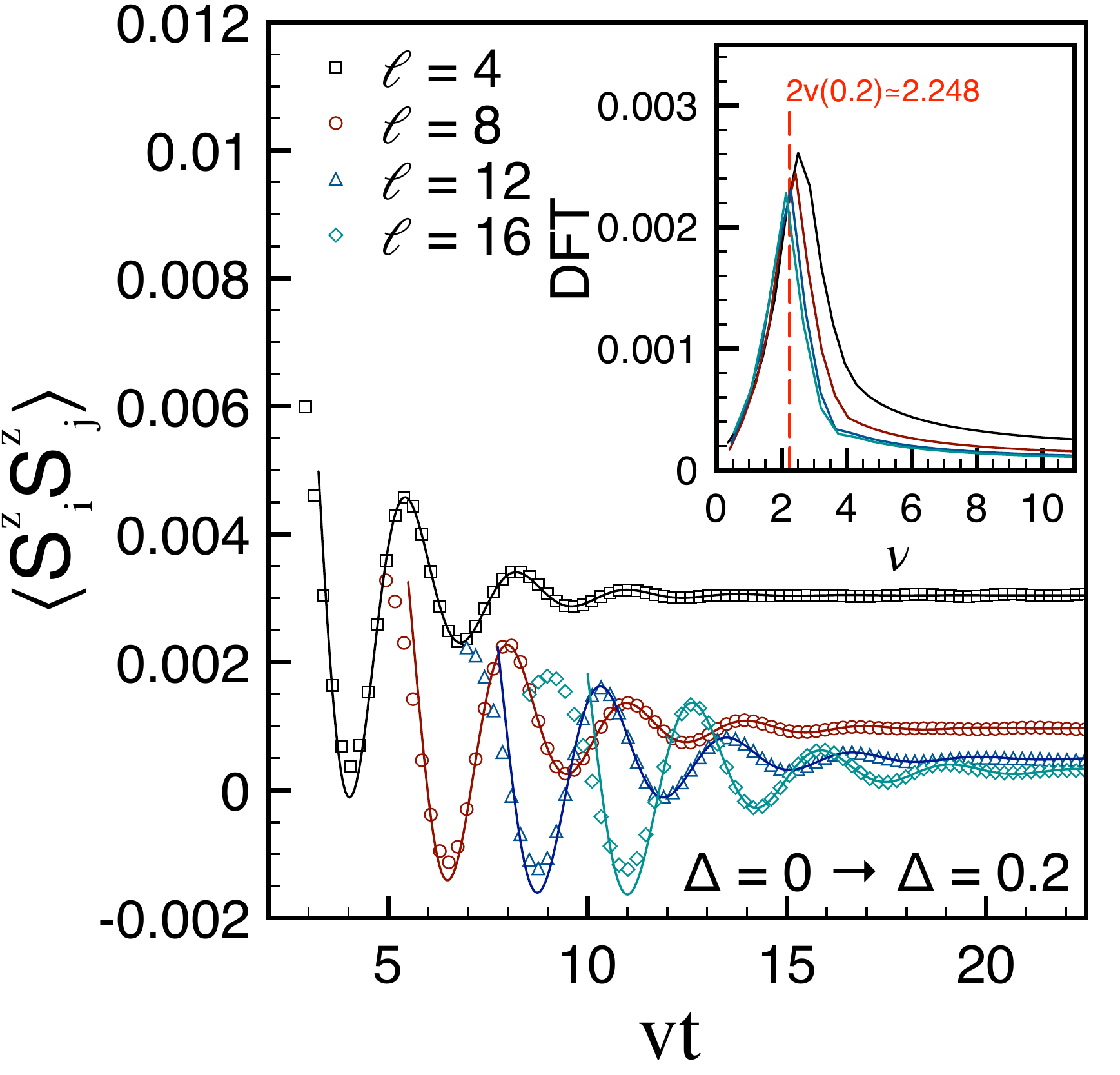}
\includegraphics[width=0.45\textwidth]{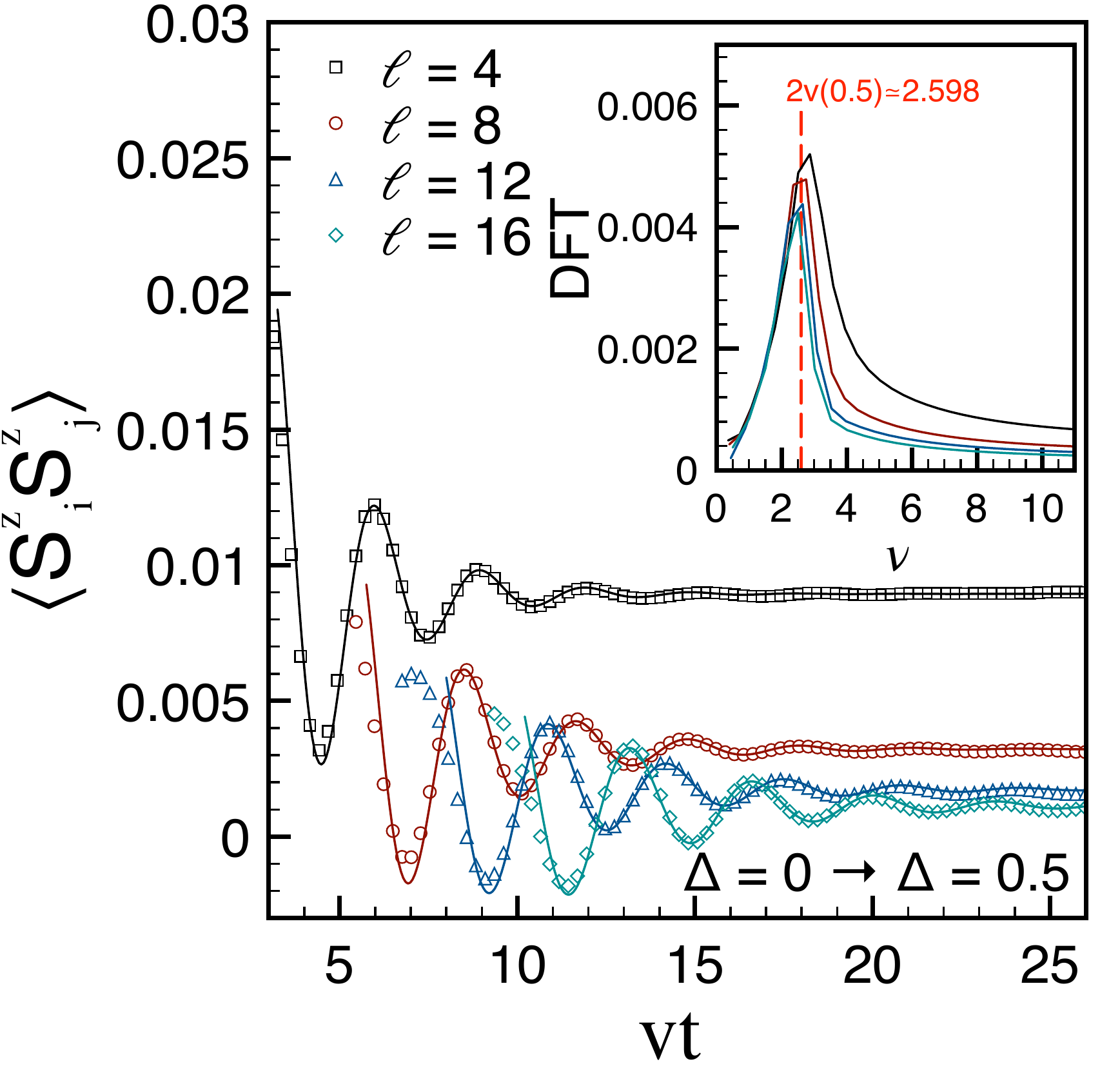}
\caption{\label{figSzSz_vs_t_long_oscill}
Large-time behaviour of $\langle S^{z}_{i}S^{z}_{j}\rangle$ with $\ell=j-i$.
The data (symbols) are compared with the best fit of Eq. (\ref{heur}). 
In the inset we report the DFT  of $\langle S^{z}_{i}S^{z}_{j}\rangle$, 
which displays a peak close to $\nu_0=2v$.} 
\end{figure}

%%%%%%%%%%%%%%%%%%%%%%%%%%%%%%%%%%%%%%%%
We have seen that inside the light cone the longitudinal correlation
function displays strong oscillatory behaviour, that is not captured
by the LL approximation. We now turn to a parameterization of this
effect. We have found that the numerical data is well described by the
functional form
\be
\langle S^{z}_{j}S^{z}_{j+\ell}\rangle\approx
B+ C \exp(-D t)\sin(2v t+\phi).
 \label{heur}
\ee
Here $v$ is equal to the ground state velocity of sound and
$\{B, C,D,\phi\}$ are constants that are used to obtains best
fits of the numerical results to (\ref{heur}). 
We use the time window $[(\ell+\delta\ell)/(2v),20]$ and then vary
$\delta\ell$ in order to optimize the fit. As is shown in
Fig.~\ref{figSzSz_vs_t_long_oscill}, the quality of these fits is
quite good.
%%%%%%%%%%%% FIGURE L=64 SzSz   vs t (long) %%%%%%%%%%%%%
%%%%%%%%%%%%%%%%%%%%%%%%%%%%%%%%%%%%%%%%%%

In order to quantify how well the oscillatory behaviour can be 
described in terms of a single frequency $\nu_0=2v$ we have carried
out a Fourier analysis of our numerical data. Discrete Fourier
transforms (DFT) of $\langle S^{z}_{i}S^{z}_{j}\rangle$ are shown in
the insets of Fig. \ref{figSzSz_vs_t_long_oscill}. We see that the
oscillation frequencies form a narrow band peaked around
$\nu_0=2v$. Moreover, as the distance $\ell=j-i$ increases, the the
frequency distribution narrows and its centre approaches $\nu_0$.
We also performed fits of our numerical results to (\ref{heur}) treating
$v$ as a fitting parameter. As expected we found that the best value
of $v$ is then extremely close to the sound velocity.

For quenches to $\Delta<0$, the accessible time window $t\in[0,20]$ 
is too small to assess the applicability of (\ref{heur}). The reason
is that the sound velocity $v$ decreases rapidly with decreasing
$\Delta$, cf. Table~\ref{tab1}. This in turn results in a small
oscillation frequency in (\ref{heur}), and concomitantly few
oscillations inside the light cone at times $t<20$. At the same time
our numerical computations are still limited by the growth of
entanglement entropy, which exhibits only a weak dependence on the
sign of $\Delta$. This precludes us from exploring larger time
windows. 

The fact that the oscillation frequency equals $\nu=2v$ gives us an
indication of which processes may underlie this behaviour. According to
the approach proposed in Ref.~\cite{ce-13}, the time evolution of the
spin-spin correlation function for a quantum quench from the initial
state $|\Psi_0\rangle$ is given by 
\be
\langle S^{z}_{j}S^{z}_{j+\ell}\rangle= \lim_{N\to\infty}{\rm Re}\sum_{|\psi_n\rangle}
e^{{\cal E}^*_{\Phi_s}-{\cal
    E}^*_{\psi_n}+it(\omega_{\psi_n}-\omega_{\Phi_s})}
\langle\psi_n|S^z_jS^z_{j+\ell}|\Phi_s\rangle\ .
\label{QA}
\ee
Here $|\psi_n\rangle$ are eigenstates of the Hamiltonian and
$\omega_{\psi_n}$ the corresponding energies, $|\Phi_s\rangle$ is a
particular simultaneous eigenstates of local conservation laws of the
spin-1/2 XXZ chain that describes the steady state, $\omega_{\Phi_s}$
its energy, and ${\cal E}_\psi\equiv-\ln\langle\psi|\Psi_0\rangle$.
It was argued in Ref.~\cite{ce-13} that only states with finite energy
differences relative to $\omega_{\Phi_s}$ contribute to
(\ref{QA}). Our numerical analysis suggests that the oscillatory
behaviour inside the light cone is well characterized by a single
frequency $\nu_0=2v$. In (\ref{QA}) this corresponds to energy
eigenstates $|\psi_n\rangle$ with 
\be
\omega_{\psi_n}-\omega_{\Phi_s}=\pm 2v.
\ee
The spectrum of the XXZ chain in the vicinity of the representative
state $|\Phi_s\rangle$ can be calculated \cite{bel-14} by a
generalized thermodynamic Bethe Ansatz \cite{mc-12b}. The most
important excitations are particle-hole excitations with energy
\be
\omega_\alpha(k^p,h^h)=\epsilon_\alpha(k^p)-\epsilon_\alpha(k^h)\ ,
\ee
where the index $\alpha$ runs over the different types (``strings'')
of allowed excitations at anisotropy $\Delta$ and $\epsilon_\alpha(k)$
is their dressed energy. We conjecture (cf. Ref.~\cite{bel-14}) that for the
quenches considered here, the bandwidth of $\epsilon_1(k)$ is
approximately equal to that of the equilibrium spinon dispersion
$\epsilon_s(k)$, and that the extrema of $\epsilon_1(k)$ occur at
$k=0,\pi/2,\pi$ (we are working at zero magnetization). For the zero
temperature equilibrium spinon dispersion we have \cite{book-ba}
\be
|\epsilon_s(\pi/2)-\epsilon_s(0)|=2v.
\ee
These considerations suggest that the oscillatory behaviour inside the
light cone originates in particle-hole excitations above the
representative state connecting saddle points of $\epsilon_1(k)$ at
$k=\pi/2$ and $k=0$. These are high-energy degrees of freedom manifestly
beyond the applicability of the Luttinger liquid approximation.
 
%%%%%%%%%%%%%%%%%%%%%%%%%%%%%%%%%%%%%%%%%%
\subsection{Stationary behaviour} 
\begin{figure}[ht]
\includegraphics[width=0.48\textwidth]{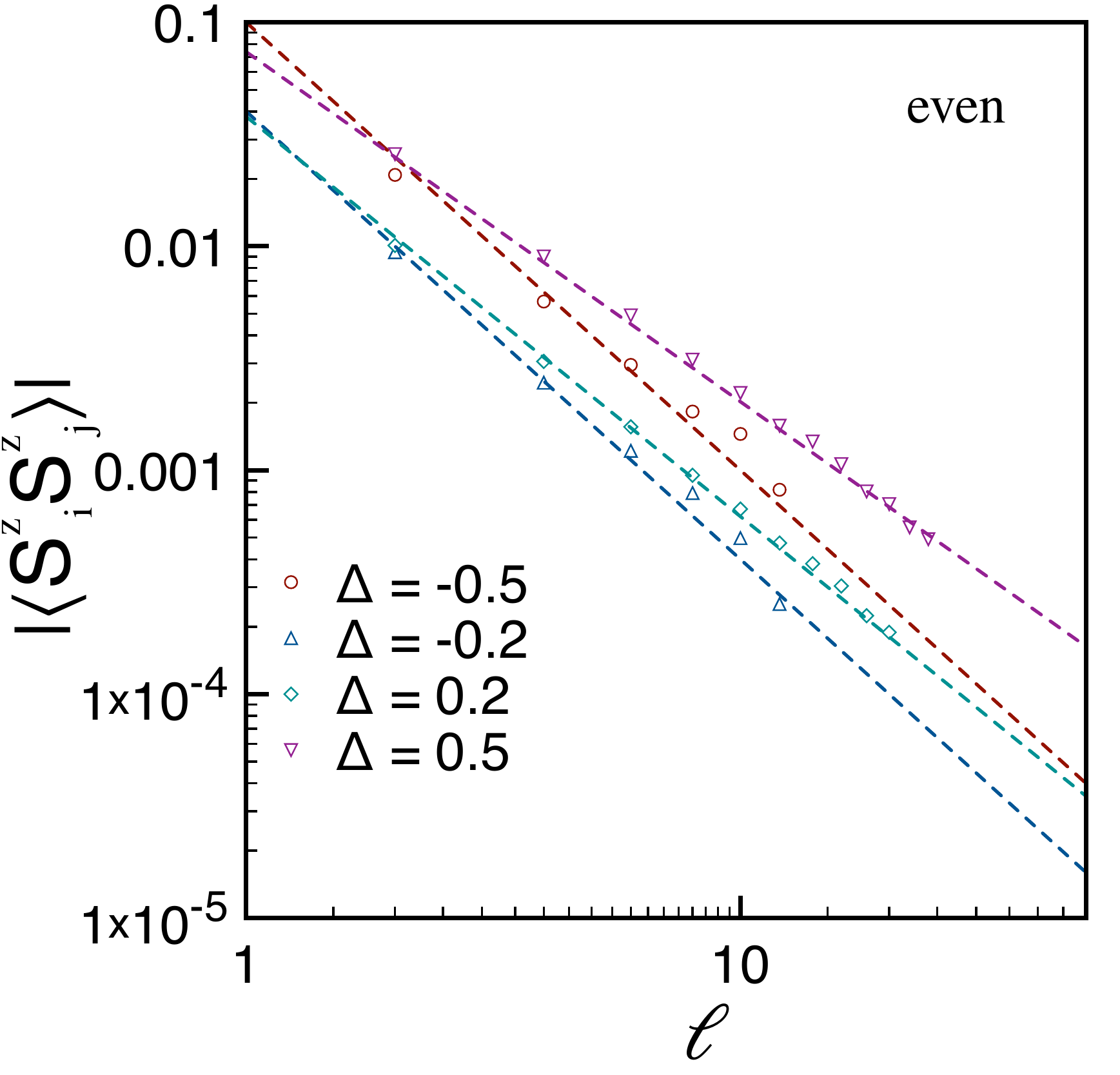}
\includegraphics[width=0.48\textwidth]{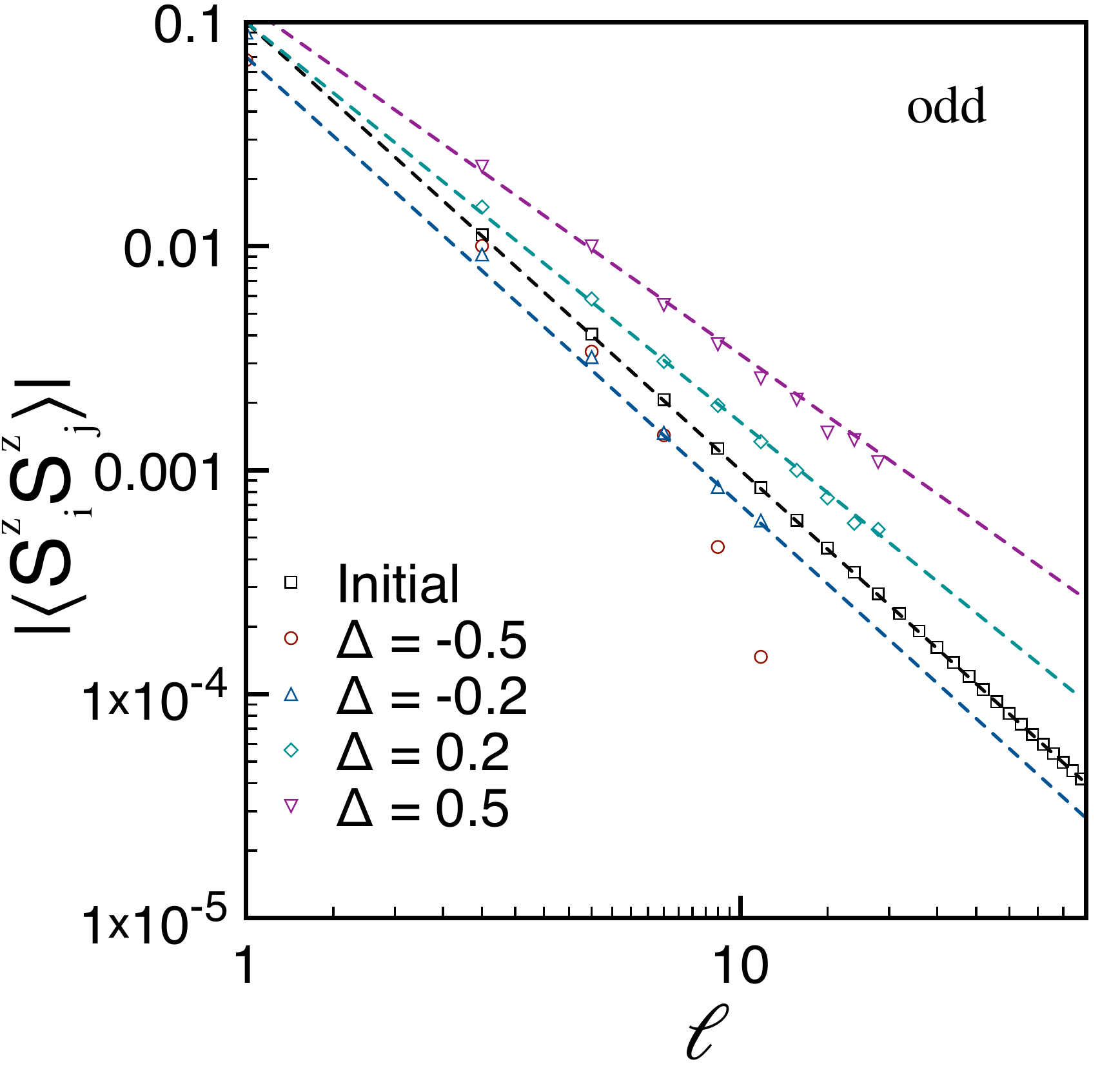}
\caption{\label{figSzSz_vs_x} 
Longitudinal correlation function $\langle S^{z}_{i}S^{z}_{j}\rangle$  
at time $t = 20$ for several different interaction strengths $\Delta$
as a function of the distance $\ell=j-i$, which is taken to be even in
the left panel and odd in the right one. The dashed lines are the
asymptotic LL predictions (\ref{LLpred}). In the right panel the
initial ($t=0$) correlations are shown as back squares. The initial
correlations for even $\ell$ are equal to zero.
}
\end{figure}

%%%%%%%%%%%%%%%%%%%%%%%%%%%%%%%%%%%%%%%%%%
Finally, we turn to the late time behaviour of the longitudinal
correlations. The latest time accessible by our numerical computations
is $t=20$. In Fig. \ref{figSzSz_vs_x} we display the dependence of
$|\langle S^z_jS^z_{j+\ell}\rangle|$ on the separation $\ell$ at that time.
As the correlator displays even/odd effects due to the presence of
both smooth and staggered contributions we show the results for even
and odd values of $\ell$ in separate graphs. The Luttinger liquid
approximation (\ref{Eq_LM_SzSz}) predicts that at late times
\begin{equation}
\lim_{t\to\infty}|\langle S^z_jS^z_{j+\ell}\rangle|\sim
\cases{
\ell^{-2} & for $\Delta<0$\ ,\\
\ell^{-\beta-2} & for $\Delta>0$.}
\label{LLpred}
\end{equation}
where the exponent $\beta$ is given in Table \ref{tab1}.
Although our numerical results at $t=20$ exhibit small oscillations
(probably non-stationary), they agree well with the asymptotic LL
prediction for all values of $\Delta$ except for $\Delta=-0.5$ and odd
$\ell$. This disagreement is not surprising, because for this value of
$\Delta$ the velocity $v$ is small and the correlations at $t=20$ are
therefore not yet stationary for the largest shown values of $\ell$.
 %%%%%%%%%%%% FIGURE  SzSz   vs x %%%%%%%%%%%%%
%%%%%%%%%%%%%%%%%%%%%%%%%%%%%%%%%%%%%%%%%%

% 
%Nevertheless, at least for $\Delta < 0$ and ($\ell \lesssim 20$), it is evident how the initial behavior 
%($\sim \ell^{-2}$) relaxes, after the information has been propagated along the lattice, toward the
%leading asymptotic stationary behavior, which for positive interactions it is given by $\ell^{-\beta-2}$, 
%i.e. the smooth contribution in the LM prediction.
%For $\Delta = -0.2$ such a behavior seems still to be dominant; however, already for $\Delta=-0.5$, 
%even though we do not have many stationary data (and therefore our conclusion cannot be  definitive),
%the correlation function seems to show a stationary behavior closer to $\ell^{-2}$, i.e.
%the staggered contribution in the LM prediction.

%%%%%%%%%%%%%%%%%%%%%%%   
\section{Conclusions} 
\label{concl} 
%%%%%%%%%%%%%%%%%%%%%%%

We have considered interaction quenches from the ground state of the
XX model to the spin-1/2 Heisenberg XXZ chain (\ref{H_XXZ}) in the
quantum critical regime. In equilibrium the low-energy physics of the
post-quench Hamiltonian is described by Luttinger liquid theory, and
we have investigated to that extent this description applies to
dynamics out of equilibrium. We have computed two-point correlation
functions of quantum spins in the XXZ chain and compared our results
to predictions obtained by means of a Luttinger liquid approximation.
Our work extends previous studies, which focused on other observables
\cite{krs-12,cbp-13}. Our main results are as follows. We found that
on the time scales accessible to us the transverse correlator $\langle
S^x_jS^x_{j+\ell}\rangle$ is well described by LL theory. Small
deviations are seen in the vicinity of the light cone,
i.e. $2vt\approx\ell$. The good agreement between the LL approximation
and the numerical results is surprising and surpasses our most optimistic
expectations.  
Conversely, LL theory fails to give a good description of the
longitudinal two-point function $\langle S^z_jS^z_{j+\ell}\rangle$
except at rather short times. This implies that the longitudinal
correlations are more strongly affected by corrections to LL theory.

It is tempting to speculate that this difference in the applicability
of the LL approximation could be related to the locality properties of
the observables $S^{x,z}_j$ with respect to the elementary excitations
that spread the correlations. While $S^z_j$ is local in this sense,
$S^x_j$ is not. This can be seen by noting that the elementary
excitations are dressed versions of the Jordan-Wigner fermions, and
the expression of $S^x_j$ in terms of the latter involves a string
operator. It has been previously noted \cite{rsm-10,cef} that local
and non-local operators exhibit qualitatively different quench
dynamics in particular examples. If our observations are indeed
related to locality properties of observables, one would expect
analogous behaviour in other models, whose low-energy equilibrium
properties are described by Luttinger liquids. One such example
is that of interaction quenches in the Lieb-Liniger model. Here
the field correlator (one particle density matrix) would then be
well described by the LL approximation, while the density-density
correlation would not. This is a opportune conjecture because in
the Lieb-Liniger model interaction quenches (between two finite
values of the interaction) are experimentally feasible, but there is
still no analytic or numerical method to tackle its non-equilibrium
dynamics. 

Finally, our results are relevant to the physics of prethermalization
\cite{kehrein,exp-pt,kwe-11,mmgs-13,ekmr-14,bck-14,bf-15,begr-15}
in models with weak integrability breaking. A prethermalized regime
has been observed in experiments \cite{exp-pt} and described using
Luttinger liquid methods. Our work shows that on intermediate time
scales genuine high-energy effects need to be taken into account as
well as  ``irrelevant'' perturbations to the Luttinger liquid
Hamiltonian \cite{m-12}. 
 
\section*{Acknowledgments}

All authors acknowledge the financial support by the ERC under
Starting Grant  279391 EDEQS. The work of FHL was supported by the
EPSRC under grants EP/J014885/1 and EP/I032487/1.

\appendix 

%%%%%%%%%%%%%%%%%%%%%%%%%%%%%%%%%%%%%%%%%%%%%%%%%%%
\section{Interaction quench in a Luttinger liquid}
\label{app:LL}
%%%%%%%%%%%%%%%%%%%%%%%%%%%%%%%%%%%%%%%%%%%%%%%%%%%
Our starting point is the Luttinger liquid Hamiltonian
\be
{\cal H}(\Delta) =\frac{v}{2} \int dx \left[K (\partial_x\theta)^2
+\frac{1}K (\partial_x\phi)^2\right].
\label{HLL}
\ee
After carrying out a canonical transformation
\be
\label{canonical}
\tilde{\phi}=\tilde\varphi+\tilde{\bar\varphi}=
\frac{\phi}{\sqrt{K}}\ ,\quad
\tilde{\theta}=\tilde\varphi-\tilde{\bar\varphi}=\theta\sqrt{K},
\ee
the time evolution (at $t>0$) of the chiral fields is described by the
mode expansion 
\bea
\label{t>0}
\tilde{\varphi}(t,x)&=&
\tilde{Q}+\frac{\tilde{P}}{2L}(vt-x)+\sum_{n>0}\frac{1}{\sqrt{4\pi n}}
\left[\tilde{\beta}_n e^{ik_n(x-vt)}+{\rm h.c.}\right],\nonumber\\
\tilde{\bar\varphi}(t,x)&=&\tilde{\bar Q}+\frac{\tilde{\bar P}}{2L}(vt+x)+
\sum_{n>0}\frac{1}{\sqrt{4\pi    n}}
\left[\tilde{\beta}_{-n} e^{-ik_n(x+vt)}+{\rm h.c.}\right].
\eea
Here we consider a ring of circumference $L$, $k_n=2\pi n/L$, and the
various mode operators fulfil commutation relations
\be
[Q,P]=i=[\bar Q,\bar P]\ ,\quad [\tilde\beta_n,\tilde\beta^\dagger_m]=\delta_{n,m}.
\ee

At $t=0$ our system is prepared in the ground state of (\ref{HLL})
with $K=1$ and $v=Ja_0$. This state is conveniently constructed
through a mode expansion of ${\cal H}(\Delta=0)$, which reads
\bea
{\varphi}(t=0,x)&=&
{Q}-\frac{{P}}{2L}x+\sum_{n>0}\frac{1}{\sqrt{4\pi n}}
\left[\beta_n e^{ik_nx}+{\rm h.c.}\right],\nonumber\\
{\bar\varphi}(t=0,x)&=&{\bar Q}+\frac{{\bar P}}{2L}x+
\sum_{n>0}\frac{1}{\sqrt{4\pi    n}}
\left[{\beta}_{-n} e^{-ik_nx}+{\rm h.c.}\right].
\label{t=0}
\eea
The ground state at $t=0$ is then defined through the property
\be
\beta_n|0\rangle_0=0=P|0\rangle_0=\bar{P}|0\rangle_0.
\label{vac}
\ee
Next, we use the fact that the mode operators in (\ref{t=0}) and
(\ref{t>0}) are related through the transformation
(\ref{canonical}), which implies that
\bea
\tilde\beta_n&=&\frac{1}{2}\left[\frac{1}{\sqrt{K}}+\sqrt{K}\right]\beta_n
+\frac{1}{2}\left[\frac{1}{\sqrt{K}}-\sqrt{K}\right]\beta^\dagger_{-n}.
%\tilde{Q}&=&\frac{1}{2}\left[\frac{1}{\sqrt{K}}+\sqrt{K}\right]Q
%+\frac{1}{2}\left[\frac{1}{\sqrt{K}}-\sqrt{K}\right]\bar Q
\label{bogo}
\eea
The zero mode operators are related by analogous expressions. We are
now in a position to work out two point functions of local operators
after our interaction quench. Let us consider
\be
F(t,x)={}_0\langle 0|
\exp\left(-i\sqrt{\frac{\pi}{K}}\tilde\theta(t,x)\right)
\exp\left(i\sqrt{\frac{\pi}{K}}\tilde\theta(t,0)\right)
|0\rangle_0\ .
\ee
This can be calculated using the mode expansions (\ref{t>0}), and then
expressing the mode operators using (\ref{bogo}). The resulting
expression is then evaluated using
\be
{}_0\langle 0|e^{\gamma\beta_n+\delta b^\dagger_n}|0\rangle_0=e^{\gamma\delta/2},
\ee
which is a simple consequence of (\ref{vac}). As we are ultimately
interested in the thermodynamic limit we ignore the contributions
arising from the zero modes. We find
\bea\fl
\ln\left(F(t,x)\right)&=&-\sum_{n=1}^\infty\frac{1}{n}\Big[
\frac{1+K^2}{8K^2}(2-e^{ik_nx}-e^{-ik_nx})\nn\fl
&&\qquad+\frac{1-K^2}{16K^2}(e^{ik_n(x+2vt)}+e^{ik_n(x-2vt)}-2e^{2ik_nvt}+{\rm c.c})\Big].
\eea
In order to proceed, we regulate the remaining sum by multiplying the
summand by a factor $e^{-2\pi\epsilon n}$. For $\epsilon>0$ the sum then
exists and is evaluated using
\be
\ln(1-z)=-\sum_{n=1}^\infty \frac{z^n}{n}.
\ee
After a short calculation we arrive at
\be
F(t,x)=\left[1+\frac{x^2}{a^2}\right]^{-\frac{1+K^2}{8K^2}}
\left[\frac{[(x+2vt)^2+a^2][(x-2vt)^2+a^2]}{
[(2vt)^2+a^2]^2}\right]^{\frac{1-K^2}{16K^2}},
\label{cutoffF}
\ee
where $a=\epsilon L$ is a short distance cutoff. In order to make
contact with the XXZ spin chain, we need to set $a$  proportional to the
lattice spacing $a_0$ (because the cutoffs employed in the initial and final Hamiltonian are not the same), 
use that $|x|, |x\pm vt|, vt \gg a_0$, and finally employ
(\ref{S-bos}). This gives the result (\ref{Eq_LM_SxSx}) quoted in the
main text. All other correlators are evaluated in an analogous fashion.

%%%%%%%%%%%%%%%%%%%%%%%%%%%%%%%%%%%%%%%%%%%%%%%%%%%%%%%%%%%%

 \end{document}